\documentclass[11pt,a4paper]{article}

\usepackage{amsmath,amssymb,tikz,hyperref,empheq}
\usepackage{graphicx}
 \allowdisplaybreaks
\def\be{\begin{equation}}
\def\ee{\end{equation}}
\def\ba#1\ea{\begin{align}#1\end{align}}
\def\bg#1\eg{\begin{gather}#1\end{gather}}
\def\bm#1\em{\begin{multline}#1\end{multline}}
\def\bmd#1\emd{\begin{multlined}#1\end{multlined}}

\def\a{\alpha}

\def\e{\epsilon}

\def\({\left(}
\def\){\right)}
\def\[{\left[}
\def\]{\right]}

\def \be {\begin{equation}}
\def \ee {\end{equation}}
\def \ba {\begin{array}}
\def \ea {\end{array}}
\def \bea{\begin{eqnarray}}
\def \eea{\end{eqnarray}}

\def \a {\alpha}

\def \e {\epsilon}

\def\bea{\begin{eqnarray}}
\def\eea{\end{eqnarray}}

\newcommand{\bit}{\begin{itemize}}  \newcommand{\eit}{\end{itemize}}
\newcommand{\ben}{\begin{enumerate}}  \newcommand{\een}{\end{enumerate}}

\long\def\symbolfootnote[#1]#2{\begingroup%
\def\thefootnote{\fnsymbol{footnote}}\footnote[#1]{#2}\endgroup}

\newcommand{\sysu}{{\it School of Physics and Astronomy, Sun Yat-Sen University, 2 Daxue Road, Zhuhai 519082, China}}

\begin{document}
\thispagestyle{empty}
\begin{center}

~\vspace{20pt}

{\Large\bf AdS/BCFT and Island for curvature-squared gravity }

\vspace{25pt}

Qi-Lin Hu, Dongqi Li, Rong-Xin Miao ${}$\symbolfootnote[1]{Email:~\sf
  miaorx@mail.sysu.edu.cn} and Yu-Qian Zeng

\vspace{10pt}${}$\sysu

\vspace{2cm}

\begin{abstract}
In this paper, we investigate AdS/BCFT for curvature-squared gravity.  To warm up, we start with Gauss-Bonnet gravity. We derive the one point function of stress tensor and show that the central charge related to the norm of displacement operator is positive for the couplings obeying causality constraints.  Furthermore, by imposing the null energy condition on the end-of-the-world brane, we prove the holographic g-theorem for Gauss-Bonnet gravity. This corrects a wrong point of view in the literature, which claims that the holographic g-theorem is violated for Gauss-Bonnet gravity. As a by-product, we obtain the boundary entropy and A-type boundary central charges in general dimensions. We also study AdS/BCFT for general curvature-squared gravity. We find that it is too restrictive for the shape of the brane and the dual BCFT is trivial if one imposes Neumann boundary conditions for all of the gravitational modes. Instead, we propose to impose Dirichlet boundary condition for the massive graviton, while imposing Neumann boundary condition for the massless graviton. In this way, we obtain non-trivial shape dependence of stress tensor and well-defined central charges. In particular, the holographic g-theorem is satisfied by general curvature-squared gravity.  Finally, we discuss the island and show that the Page curve can be recovered for Gauss-Bonnet gravity.  Interestingly, there are zeroth-order phase transitions for the Page curve within one range of couplings obeying causality constraints.  Generalizing the discussions to holographic entanglement entropy and holographic complexity in AdS/CFT, we get new constraints for the Gauss-Bonnet coupling,  which is stronger than the causality constraint.
\end{abstract}

\end{center}

\newpage
\setcounter{footnote}{0}
\setcounter{page}{1}

\tableofcontents

\section{Introduction}

Recently, a great breakthrough towards the resolution of the black hole information paradox has been made \cite{Penington:2019npb,Almheiri:2019psf}, where double holography and island play an important role. See
\cite{Almheiri:2020cfm,Almheiri:2019hni,Rozali:2019day, Chen:2019uhq,Almheiri:2019yqk,Almheiri:2019psy,Kusuki:2019hcg,
   Balasubramanian:2020hfs,Geng:2020qvw,Chen:2020uac,Ling:2020laa,
   Kawabata:2021hac,Bhattacharya:2021jrn,Kawabata:2021vyo,
Geng:2021hlu,Krishnan:2020fer,Neuenfeld:2021bsb,Chen:2020hmv,Ghosh:2021axl,Omiya:2021olc,Bhattacharya:2021nqj,Geng:2021mic,Sun:2021dfl,Chou:2021boq,Ahn:2021chg,He:2021mst,Yu:2021rfg,Wang:2021xih,Alishahiha:2020qza} for some recent works.  As a generalization of the AdS/CFT correspondence \cite{Maldacena:1997re,Gubser:1998bc,Witten:1998qj}, double holography has a close relation to brane world theory  \cite{Randall:1999ee,Randall:1999vf,Karch:2000ct} and AdS/BCFT \cite{Takayanagi:2011zk,Fujita:2011fp,Nozaki:2012qd,Miao:2018qkc,Miao:2017gyt,Chu:2017aab,Chu:2021mvq}.  Here BCFT means a conformal field theory defined on a manifold with a boundary, where suitable boundary conditions are imposed \cite{Cardy:2004hm,McAvity:1993ue}.  Recently, a novel doubly holographic model called wedge holography has been proposed \cite{Akal:2020wfl}.  Remarkably, the effective theory of wedge holography on the brane is a ghost-free higher derivative gravity, which includes a massless mode and behaves like Einstein gravity in many aspects \cite{Hu:2022lxl}.  Wedge holography can be regarded as a holographic dual of the edge mode living on the boundary (codim-1 defect) \cite{Miao:2021ual}. Generalizing wedge holography to codim-m defects, \cite{Miao:2021ual} proposes the so-called cone holography.   See also \cite{Bousso:2020kmy,Miao:2020oey,Geng:2020fxl,Uhlemann:2021nhu,Uhlemann:2021itz} for some related works on wedge/cone holography.

In this paper, we investigate AdS/BCFT for higher derivative gravity \footnote{Previous works on this topic include new massive gravity in three dimensions \cite{Kwon:2012tp} and Gauss-Bonnet gravity in higher dimensions \cite{Najian:2014waa}. In this paper, we discuss AdS/BCFT for general curvature-squared gravity and gain new insights.}. 
Higher derivative gravity is interesting in many aspects. First,  the CFT/BCFT dual to Einstein gravity is quite special, whose central charges are not independent. Take 4d CFT/BCFT as an example, the bulk central charges are the same for Einstein gravity, i.e., $a=c$.  Considering higher derivative gravity can make the central charges different, i.e., $a\ne c$, so that one can study more general classes of CFT/BCFTs.  Of course, Gauss-Bonnet gravity can do the same job for 4d CFT/BCFT. However, for 6d and higher dimensional CFT/BCFTs, Gauss-Bonnet gravity/Lovelock gravity cannot make all the central charges independent. Thus, it is necessary to consider general higher curvature gravity in order to cover more general classes of CFT/BCFTs.  Second, string theory predicts higher derivative corrections in the gravitational action. Third, due to the loop corrections, the higher derivative terms naturally appear in the effective theory of gravity.  Fourth, maybe most interestingly, the general higher curvature gravity is renormalizable \cite{Stelle}. Although it may suffer the ghost problem, the ghost-free and potentially renormalizable higher derivative gravity can be constructed by choosing carefully the parameters of the theory \cite{Lu:2011zk,Biswas:2011ar,Modesto:2017sdr}.  Thus, it is interesting to study higher derivative gravity in AdS/BCFT.

\begin{figure}[t]
\centering
\includegraphics[width=10cm]{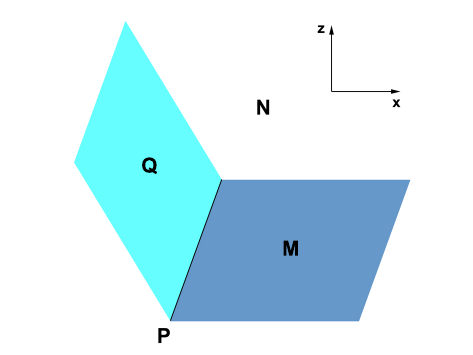}
\caption{Geometry of AdS/BCFT, where $M$ is a d-dimensional manifold with a boundary $P=\partial M$, $N$ is (d+1)-dimensional AdS space bounded by $M$ and $Q$, i.e., $\partial N=M\cup Q$. AdS/BCFT proposes that the gravity theory in the bulk $N$ is dual to the BCFT on the AdS boundary $M$. }
\label{AdSBCFT}
\end{figure}

The geometry of AdS/BCFT is shown in Fig.\ref{AdSBCFT} , where $M$ is a $d$-dimensional manifold where BCFT lives, $P=\partial M$ is the boundary of $M$, $Q$ is the $d$-dimensional end-of-the-world brane,
and $N$ is  $(d+1)$-dimensional AdS space in the bulk, which is bounded by $M$ and $Q$, i.e., $\partial N=M\cup Q$. The key problem of holographic BCFT is to determine the location of the end-of-the-world brane $Q$. To do so, we need a well-defined Gibbons-Hawking-York (GHY) boundary term and suitable boundary conditions. Unfortunately, the well-defined GHY boundary term is still lacking for general higher derivative gravity. There is a famous proposal of GHY-like boundary terms for higher derivative gravity \cite{Deruelle:2009zk}. However, this proposal does not agree with the boundary terms of Gauss-Bonnet gravity and Lovelock gravity \cite{Myers:1987yn}.  Besides, as we will show in sect.3.1, this GHY-like boundary term together with Neumann boundary condition (NBC) make vanishing the extrinsic curvature on the brane generally
\begin{eqnarray}\label{Kij}
K_{\mu\nu}=0.
\end{eqnarray}
It should be mentioned that the junction condition of general higher derivative gravity also leads to (\ref{Kij}) \cite{Reina:2015gxa,Berezin:2020mas,Chu:2021uec}. This is however too restrictive, which means that the brane $Q$ must be perpendicular to the AdS boundary $M$.  As a result, the boundary entropy and the A-type boundary central charge, which characterize the boundary degrees of freedom, vanish for the dual BCFTs. 

In this paper, we try to resolve the overly restrictive problem (\ref{Kij}) for higher derivative gravity. For simplicity, we focus on curvature-squared gravity. We propose to add the standard boundary term \cite{Myers:1987yn} and impose NBC for the Gauss-Bonnet term $R_{\mu\nu\alpha\beta}R^{\mu\nu\alpha\beta}-4R_{\mu\nu}R^{\mu\nu}+R^2$, while add the GHY-like boundary term \cite{Deruelle:2009zk} and impose Dirichlet boundary condition (DBC) for $R_{\mu\nu}R^{\mu\nu}$ and $R^2$.  In this way, we relax the constraint on the shape of the brane and obtain a non-trivial holographic dual of BCFT for higher curvature gravity, where the extrinsic curvature $K_{\mu\nu}$, the boundary entropy and the A-type boundary central charge can be non-zero. 

Let us summarize our main results. We start with AdS/BCFT for Gauss-Bonnet gravity, which has a well-defined GHY boundary term without the constraint (\ref{Kij}). We derive the shape dependence of one point function of stress tensor and show that the B-type boundary central charge, which is related to the norm of displacement operator, is positive. Furthermore, by imposing the null energy condition on the brane, we prove the holographic g-theorem for Gauss-Bonnet gravity. 
We verify that our g-function can produce the boundary entropy of a half ball and the A-type boundary central charges in general dimensions. 
These are strong supports for our results.  For general curvature-squared gravity, we propose to impose DBC for the massive gravitons, while imposing NBC for the massless gravitons. In this way, we relax the constraint on the shape of the brane and obtain non-trivial boundary central charges. In particular, the holographic g-theorem is satisfied for the gravitational couplings obeying causality constraints.  Finally, we discuss the island and Page curve for Gauss-Bonnet gravity. Remarkably, the Page curve can be discontinuous and there are 
zeroth-order phase transitions of entanglement entropy within one range of couplings obeying causality constraints.  Generalizing the discussions to holographic entanglement entropy and holographic complexity in AdS/CFT, we get a new constraint  for the Gauss-Bonnet coupling,  which is stronger than the causality constraint.

The paper is organized as follows. 
In section 2, we study AdS/BCFT for Gauss-Bonnet gravity and discuss the one point function and the holographic g-theorem. In section 3, we formulate AdS/BCFT for curvature-squared gravity and resolve the overly restrictive problem. In section 4, we investigate the island and Page curve for Gauss-Bonnet gravity. Finally, we conclude with some open problems in section 5.

\section{AdS/BCFT for Gauss-Bonnet gravity}

To warm up, let us first study AdS/BCFT for Gauss-Bonnet gravity, which has a well-defined GHY boundary term.  We derive the shape dependence of one point function of stress tensor and verify that the central charge related to the norm of displacement operator is non-negative. Furthermore, we prove the holographic g-theorem for Gauss-Bonnet gravity \footnote{In \cite{Najian:2014waa}, it claims that the holographic g-theorem is violated for Gauss-Bonnet gravity. We notice that  \cite{Najian:2014waa} gets the wrong A-type boundary central charge, which results in wrong conclusions.}. These are strong supports for our results. 

The action of Gauss-Bonnet gravity is given by \cite{Buchel:2009sk}
\begin{eqnarray}\label{GBactionbulk}
  I_{\text{GB\ bulk}}=\frac{1}{16\pi G_N}\int_N d^{d+1}x\sqrt{|g|} \Big(R+\frac{d(d-1)}{L^2}+\frac{L^2 \lambda_{\text{GB}}}{(d-2)(d-3)} \mathcal{L}_{\text{GB}} \Big),\end{eqnarray}
 where $\mathcal{L}_{\text{GB}} =R_{\mu\nu\alpha\beta}R^{\mu\nu\alpha\beta}-4R_{\mu\nu}R^{\mu\nu}+R^2$ and $\lambda_{\text{GB}}$ is the parameter of GB gravity, which obeys the following constraint
  \begin{eqnarray}\label{GBconstraint}
-\frac{(d-2) (3 d+2)}{4 (d+2)^2}\le \lambda_{\text{GB}}\le \frac{(d-3) (d-2) \left(d^2-d+6\right)}{4 \left(d^2-3 d+6\right)^2},
  \end{eqnarray} 
  in order that no negative energy fluxes appear \cite{Buchel:2009sk}. We call it causality constraint in this paper.  To have a well-defined action principle, one should add suitable Gibbons-Hawking-York boundary term on the boundary. For Gauss-Bonnet gravity, it is given by \cite{Myers:1987yn}
  \begin{eqnarray}\label{GBactionbdy}
  I_{\text{GB\ bdy}}=\frac{1}{8\pi G_N} \int_{Q} d^{d}y\sqrt{|h|} \Big( K-T+ \frac{2L^2 \lambda_{\text{GB}}}{(d-2)(d-3)} (J-2 G^{ij}_{Q} K_{ij}) \Big),
  \end{eqnarray}
where $G^{ij}_{Q} $ is the intrinsic Einstein tensor on the bulk boundary $Q$, and  $J$ is the trace of 
 \begin{eqnarray}\label{Jij}
J_{ij}=\frac{1}{3}\left(2 K K_{ik}K^k_j-2 K_{ik}K^{kl}K_{lj}+K_{ij}\left(K_{kl}K^{kl}-K^2\right) \right). 
  \end{eqnarray}
  
  Taking the variation of the total action $I_{\text{GB}}=I_{\text{GB \ bulk}}+ I_{\text{GB\ bdy}}$ and focusing on the boundary terms, we get
   \begin{eqnarray}\label{variationGB}
\delta I_{\text{GB}}=\frac{1}{2} \int_{Q} d^{d}y\sqrt{|h|} T^{ij}_{\text{GB}} \delta h_{ij},
  \end{eqnarray}
where $T^{ij}_{\text{GB}}$ is the Brown-York stress tensor for Gauss-Bonnet gravity
 \begin{eqnarray}\label{TijGB}
-8\pi G_N T^{ij}_{\text{GB}}=K^{ij}-(K-T) h^{ij}+\frac{2L^2 \lambda_{\text{GB}}}{(d-2)(d-3)} (Q^{ij}-\frac{1}{3}Q h^{ij}),
  \end{eqnarray}
 and $Q$ is the trace of
   \begin{eqnarray}\label{Qij}
Q_{ij}=3J_{ij}+2 K R_{Qij}+R_{Q} K_{ij}-2K^{kl} R_{Qkilj}-4R_{Q k(i}K^k_{j)}.
  \end{eqnarray}
  To have a well-defined action principle $\delta I=0$, one can impose various boundary conditions \cite{Witten:2018lgb,Anderson:2006lqb,Anderson:2007jpe,Anderson:2010ph,York:1972sj,Papadimitriou:2005ii}. 
  For simplicity, we focus on Neumann boundary condition (NBC) on the end-of-the-world brane in this section
\begin{eqnarray}\label{NBCGB}
\text{NBC}:\ K^{ij}-(K-T) h^{ij}+\frac{2L^2 \lambda_{\text{GB}}}{(d-2)(d-3)} (Q^{ij}-\frac{1}{3}Q h^{ij})=0.
\end{eqnarray}
We leave the discussions of Dirichlet boundary condition (DBC) \cite{Miao:2018qkc} and the conformal boundary condition (CBC) \cite{Chu:2021mvq} to future works. 

It is convenient to rewrite the action (\ref{GBactionbulk}) by applying the background-field method \cite{Miao:2013nfa}. This method is quite useful for the study of Weyl anomaly \cite{Miao:2013nfa}, correlation functions \cite{Sen:2014nfa}, and entanglement/R\'enyi entropy \cite{Miao:2015iba,Miao:2015dua,Chu:2016tps} for higher derivative gravity.  Expanding $I_{\text{GB}}$ (\ref{GBactionbulk},\ref{GBactionbdy}) in terms of the background curvature 
$\bar{R}$ defined below
\begin{eqnarray}\label{background curvature1}
&&R=\bar{R}-\frac{d(d+1)}{l^2},\\ \label{background curvature2}
&&R_{\mu\nu}=\bar{R}_{\mu\nu}-\frac{d}{l^2}g_{\mu\nu},\\ \label{background curvature3}
&&R_{\mu\nu\rho\sigma}=\bar{R}_{\mu\nu\rho\sigma}-\frac{1}{l^2}(g_{\mu\rho}g_{\nu\sigma}-g_{\mu\sigma}g_{\nu\rho}),
\end{eqnarray}
we get 
 \begin{eqnarray}\label{IGBmiao}
&&I_{\text{GB}}=\frac{1}{16\pi \bar{G}_N}\int_N d^{d+1}x\sqrt{|g|} \Big(R+\frac{d(d-1)}{l^2}+\alpha l^2\ \mathcal{L}_{\text{GB}}(\bar{R}) \Big)\nonumber\\
&&\ \ \ \ \ \ + \frac{1}{8\pi \bar{G}_N}\int_Q d^{d}y\sqrt{|h|} \Big( (1+2\alpha(d-1)(d-2))(K-T)+ 2\alpha l^2 (J-2 G^{ij}_{Q} K_{ij}) \Big),
 \end{eqnarray}
where the background curvature $\bar{R}_{\mu\nu\rho\sigma}$ vanishes for an AdS space with the radius $l$, and we have reparameterized
 \begin{eqnarray}\label{reparameter1}
&&\frac{1}{ G_N}=\frac{1+2\alpha (d-1)(d-2)}{\bar{G}_{N}}, \\ \label{reparameter2}
&&\frac{1}{ L^2}=\frac{1+\alpha (d-2)(d+1)}{1+2\alpha (d-1)(d-2) }\ \frac{1}{ l^2},\\ \label{reparameter3}
&&\frac{ \lambda_{\text{GB}}}{(d-2)(d-3)} =\frac{1+\alpha (d-2)(d+1)}{(1+2\alpha (d-1)(d-2))^2}\ \alpha.
\end{eqnarray}
In the above reparameterization, the causality constraint (\ref{GBconstraint})  and the NBC (\ref{NBCGB}) become
  \begin{eqnarray}\label{GBconstraintbar}
\frac{-1}{4 (d^2-2d-2)}\le \alpha \le \frac{1}{8},
  \end{eqnarray} 
and
\begin{eqnarray}\label{NBCGBbar}
\text{NBC}:\ \left(1+2\alpha \left(d-1\right)\left(d-2\right)\right)\left(K^{ij}-\left(K-T\right)h^{ij}\right) +2\alpha l^2(Q^{ij}-\frac{1}{3}Q h^{ij})=0,
\end{eqnarray}
respectively.  Recall that $Q_{ij}$ is given by (\ref{Qij}).  For simplicity, we set  $16\pi \bar{G}_N=l=1$
in the followings of this paper.

\subsection{Shape dependence of one point function}

In this subsection, we study the the shape dependence of the one point function of stress tensor. 
We take the following ansatz of the bulk metric and the embedding function of the brane $Q$ \cite{Miao:2017aba}
\begin{eqnarray}\label{GBmetric}
&&\text{bulk metric}:\ ds^2=\frac{dz^2+dx^2+(\delta_{ab}-2x \epsilon \bar{k}_{ab} f(\frac{z}{x}) )dy^a dy^b}{z^2}+O(\epsilon^2), \\ \label{GBQ} 
&&\text{brane Q}: \ \ \ \ \ x=-\sinh(\rho) z+ O(\epsilon^2),
\end{eqnarray}
where $ \bar{k}_{ab} $ is the traceless part of the extrinsic curvature, $\e$ denotes the order of perturbations, and $\rho$ is a constant related to the tension of the brane.  Substituting (\ref{GBmetric}) into equations of motion (EOM) derived from the action (\ref{IGBmiao}) with $16\pi \bar{G}_N=l=1$, we get at the 
first-order of $O(\e)$
\begin{eqnarray}\label{GBeq}
s(s^2+1)f''(s)-(d-1) f'(s)=0,
\end{eqnarray}
where $s=z/x$. Remarkably, (\ref{GBeq}) is exactly the same as that of Einstein gravity \cite{Miao:2017aba}. This happens because we have used the background-field method \cite{Miao:2013nfa}, which helps to simplify the results a lot. Solving (\ref{GBeq}), we obtain
\begin{eqnarray}\label{GBsolution}
f(s)=1+\alpha_d \frac{s^d \, _2F_1\left(\frac{d-1}{2},\frac{d}{2};\frac{d+2}{2};-s^2\right)}{d[1+4(d-2)\alpha] },
\end{eqnarray}
where we have used the DBC $f(0)=1$ on the AdS boundary $s=z/x=0$ \cite{Miao:2017aba}.  
Note that, in the above derivations, we have assumed $s=z/x\ge 0$ (equivalently $x\ge 0$) so that $\sqrt{z^2/x^2}=z/x$ instead of $-z/x$. For one given $d$, the above function $f(z/x)$ is discontinuous at $x=0$. Thus, suitable analytic continuation must be performed in order to get a smooth function $f(z/x)$ for $-\sinh(\rho) z\le x < \infty$. The trick is as follows: we first assume $x>0$ and simplify (\ref{GBsolution}) for any given $d$, and then analytically extend the result to $x<0$. Please see \cite{Miao:2017aba} for some examples.

Imposing NBC (\ref{NBCGBbar}) on the brane (\ref{GBQ}), at the leading order $O(\e^0)$, we determine the brane tension 
\begin{eqnarray}\label{GBtension}
T=\frac{(d-1) \tanh (\rho ) \text{sech}^2(\rho ) ((4 \alpha  (d-2) d+3) \cosh (2 \rho )-4 \alpha  (d-6) (d-2)+3)}{6+12 \alpha  (d-2) (d-1)}.
\end{eqnarray}
At the sub-leading order $O(\e)$, we fix the integral constant
\begin{eqnarray}\label{alphadGB}
\alpha_d=-[1+4(d-2)\alpha] \frac{d \cosh ^d(\rho ) \left((4 \alpha  (d-2)+1) \coth ^2(\rho )+4 \alpha  (d-3) (d-2)\right)}{d (4 \alpha  (d-2)+1) \cosh ^2(\rho ) \coth ^3(\rho )+(-\coth (\rho ))^d  G},
\end{eqnarray}
where 
\begin{eqnarray}\label{alphadGBG}
G=\left((4 \alpha  (d-2)+1) \coth ^2(\rho )+4 \alpha  (d-3) (d-2)\right) \, _2F_1\left(\frac{d-1}{2},\frac{d}{2};\frac{d+2}{2};-\text{csch}^2(\rho )\right).\nonumber\\
\end{eqnarray}
According to \cite{Miao:2017aba}, a suitable analytic continuation of the hypergeometric function should be taken in order to get the smooth integral constant (\ref{alphadGB}) at $\rho=0$.  The reasons are as follows. Recall that, before the analytic continuation, the function (\ref{GBsolution}) is well-defined only for $z/x \ge 0$. As a result the integral constant (\ref{alphadGB}) is well-defined only for $z/x=-1/\sinh(\rho) \ge 0$, or equivalently, $\rho\le 0$.  One can check that (\ref{alphadGB}) is discontinuous at $\rho=0$ for any given $d$.  Thus suitable analytic continuation of (\ref{alphadGB}) much be performed in order to get a smooth function of $\alpha_d(\rho)$. The trick is that we first simplify $\alpha_d$ (\ref{alphadGB}) under the assumption $\rho<0$, and then analytically extend the result to $\rho>0$.  In this way, we obtain for $d=4$ and $d=5$,
\begin{eqnarray}\label{alphadGB4d}
\alpha_4=\frac{(1+8\alpha) \sinh (\rho ) \cosh (\rho ) \left((8 \alpha +1) \coth ^2(\rho )+8 \alpha \right)}{(\coth (\rho )+1) (4 \alpha  \sinh (2 \rho )+(12 \alpha +1) \cosh (2 \rho )+4 \alpha +1)},
\end{eqnarray}
and 
\begin{eqnarray}\label{alphadGB5d}
\alpha_5=\frac{8 (1+12\alpha) \cosh ^2(\rho ) ((36 \alpha +1) \cosh (2 \rho )-12 \alpha +1)}{6 (4 \alpha +1) \sinh (\rho )+6 (36 \alpha +1) \sinh (3 \rho )+H},
\end{eqnarray}
where 
\begin{eqnarray}\label{alphadGBH}
H=12 \cosh ^2(\rho ) ((36 \alpha +1) \cosh (2 \rho )-12 \alpha +1) \left(2 \tan ^{-1}\left(\frac{\sinh (\rho )}{\cosh (\rho )+1}\right)+\frac{\pi }{2}\right).
\end{eqnarray}
We verify that (\ref{alphadGB4d},\ref{alphadGB5d}) agree with the results of \cite{Miao:2017aba,Chalabi:2021jud} for Einstein gravity. This can be regarded as a test of our results.

\begin{figure}[t]
\centering
\includegraphics[width=10cm]{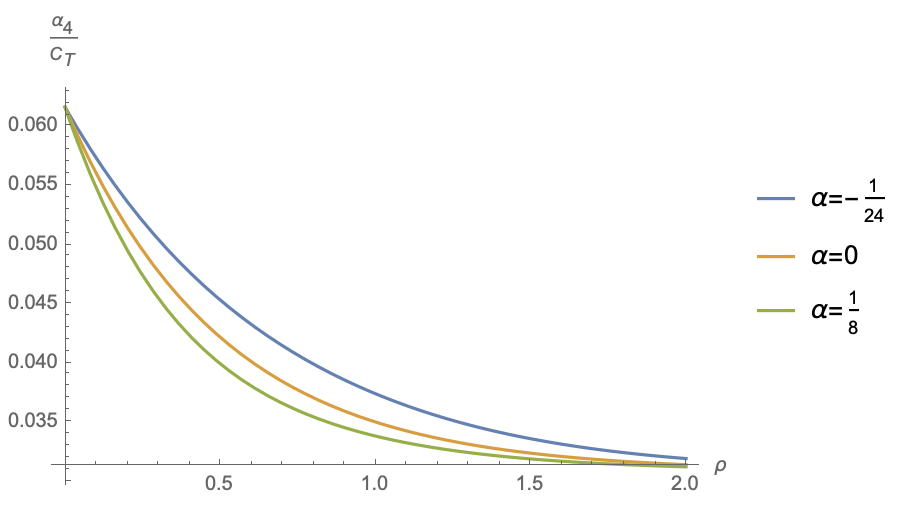}
\includegraphics[width=10cm]{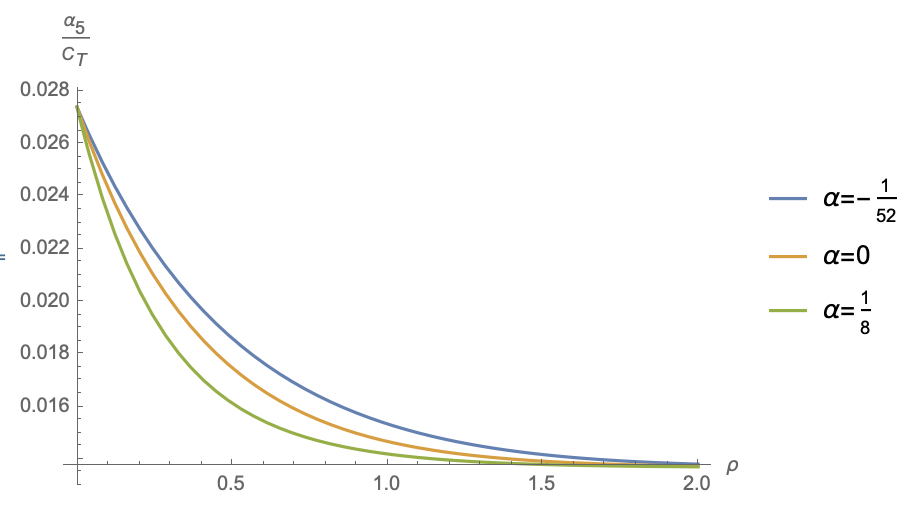}
\caption{Casimir coefficient $\a_d$ for $d=4$ (above) and $d=5$ (below). The blue line, orange line and green line correspond to the lower bound, zero and the upper bound of the Gauss-Bonnet coupling $\a$. The Casimir coefficient $\a_d$ is positive and decreases with the brane tension $\rho$. The smaller $\alpha$ is, the larger  $\alpha_d/C_T $ is, where $C_T$ is the central charge defined by (\ref{twopointTij}). }
\label{adGB}
\end{figure}

Now we are ready to derive the one point function of stress tensor. According to \cite{Sen:2014nfa}, the holographic stress tensor of GB gravity (\ref{IGBmiao}) is given by \footnote{In fact, the holographic stress tensor for general higher curvature gravity such as (\ref{Ibulkcurvature}) is still given by (\ref{holoTijGB}) \cite{Sen:2014nfa}. }
\begin{eqnarray}\label{holoTijGB}
T_{ij}=d[1+4(d-2)\alpha] h^{(d)}_{ij},
\end{eqnarray}
where $h^{(d)}_{ij}$ is defined in the Fefferman-Graham expansion of the asymptotically AdS metric
\begin{eqnarray}\label{FG}
ds^2=\frac{dz^2+(g^{(0)}_{ij}+z^2g^{(1)}_{ij}+...+z^dh^{(d)}_{ij}+...) dy^i dy^j}{z^2}.
\end{eqnarray}
From (\ref{GBmetric},\ref{GBsolution},\ref{holoTijGB},\ref{FG}), we obtain the holographic stress tensor
\begin{eqnarray}\label{TijGB1}
T_{ab}=-2 \e \alpha_d \frac{\bar{k}_{ab}}{x^{d-1}}+O(\e^2),
\end{eqnarray}
which takes the expected universal form \cite{Miao:2017aba}.  Note that the Casimir coefficient $\alpha_d $ is related to the B-type central charge of boundary Weyl anomaly \cite{Miao:2017aba} and the norm of displacement operator \cite{Miao:2018dvm,Herzog:2017kkj,Herzog:2017xha}. As a result, $\alpha_d $ (\ref{alphadGB},\ref{alphadGB4d},\ref{alphadGB5d}) must be non-negative in order to define an unitary BCFT
\begin{eqnarray}\label{adpositive}
\alpha_d  \ge 0.
\end{eqnarray}
One can check that this is indeed the case provided that the brane tension is  non-negative, i.e., $T\ge 0$ ($\rho\ge 0$), and the Gauss-Bonnet coupling obeys the causality constraint (\ref{GBconstraintbar}).  See Fig.\ref{adGB} for an example. This is a strong support for our results. 

\begin{figure}[t]
\centering
\includegraphics[width=10cm]{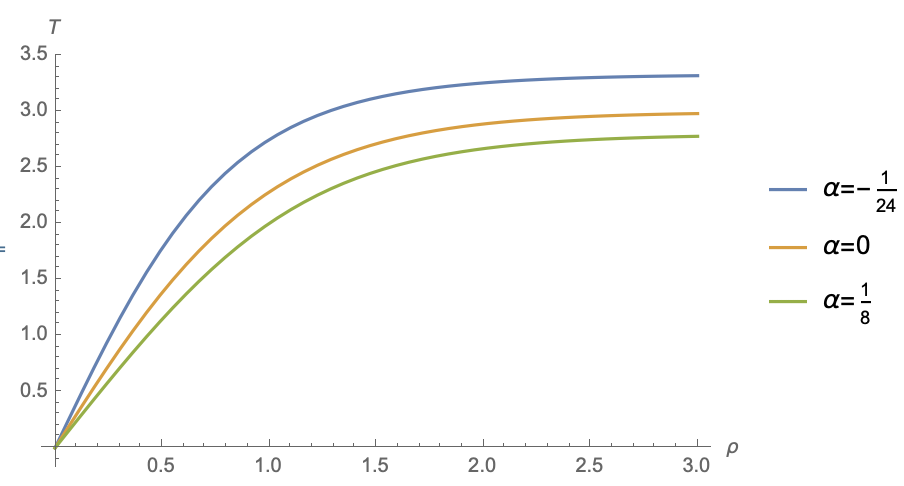}
\includegraphics[width=10cm]{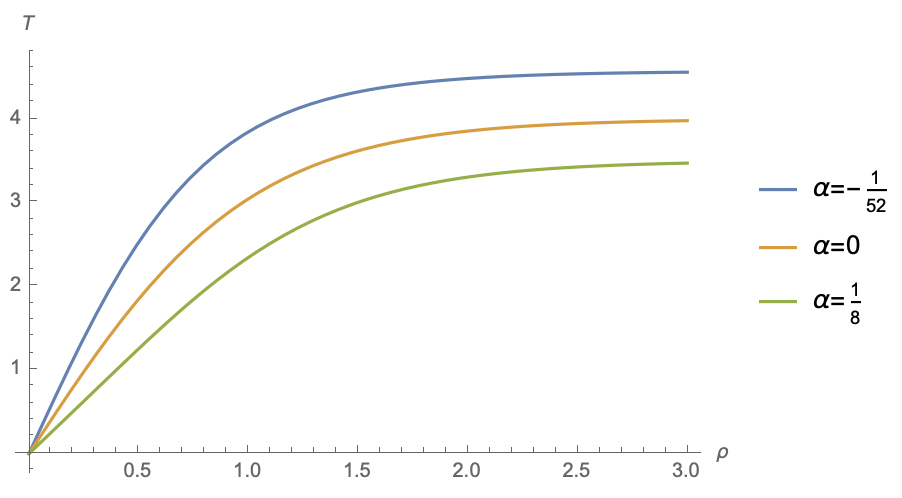}
\caption{Brane tension $T$ for $d=4$ (above) and $d=5$ (below). The blue line, orange line and green line correspond to the lower bound, zero and the upper bound of the Gauss-Bonnet coupling $\a$. The brane tension $T$ is positive for $\rho>0$ and increases with $\rho$. The smaller $\alpha$ is, the larger the tension is.}
\label{TGB}
\end{figure}

To end this subsection, let us make some comments.  First, as shown in Fig.\ref{adGB}, the Casimir coefficient $\alpha_d $ is positive and decreases with the brane tension $\rho$. The smaller $\alpha$ is, the larger  $\alpha_d/C_T $ is. Here 
\begin{eqnarray}\label{CT}
C_T=\frac{2(d+1)}{d-1}\frac{\Gamma[d+1]}{\pi^{d/2}\Gamma[d/2]}(1+4\a (d-2))
\end{eqnarray}
is the central charge defined by the two point functions of stress tensor far away from the boundary
\begin{eqnarray}\label{twopointTij}
\langle T_{ij}(x)T_{kl}(0)\rangle =\frac{C_T}{|x|^{2d}} I_{ij,kl}(x),
\end{eqnarray}
with $I_{ij,kl}$ is a dimensionless tensor.  Second, in the zero tension limit $\rho=0$, $\alpha_d/C_T $ approaches to a universal upper bound, which is independent of the Gauss-Bonnet coupling
\begin{eqnarray}\label{freeBCFTad}
\lim_{\rho\to 0} \a_d=\frac{\pi^{d/2}}{2^d(d+1) \Gamma(1+\frac{d}{2})} C_T. 
\end{eqnarray}
Remarkably, the free BCFT obeys the same relation as above  \cite{Miao:2018qkc}.
Third, in the large tension limit, $\alpha_d/C_T $  saturates the universal lower bound
\begin{eqnarray}\label{lowerboundad}
\lim_{\rho\to \infty} \a_d=\frac{\pi^{d/2}}{2^{d+1}(d+1) \Gamma(1+\frac{d}{2})} C_T,
\end{eqnarray}
which is one half of the upper bound (\ref{freeBCFTad}). Fourth, as shown in Fig. \ref{TGB}, the brane tension (\ref{GBtension}) is positive for $\rho>0$ and increases with $\rho$.  

\subsection{Holographic g-theorem}

In this subsection, we study the holographic g-theorem for Gauss-Bonnet gravity.  We propose a new g-function, which can reproduce the A-type boundary central charge and the boundary entropy of a half ball in general dimensions. To the best of our knowledge, this is the first time such a g-function is constructed in general dimensions. By imposing the null energy condition on the brane, we prove that the g-function decreases under Renormalization Group (RG) flows. 

Following \cite{Takayanagi:2011zk,Fujita:2011fp}, we focus on the AdS spacetime in the bulk \footnote{Note that we are studying the holographic g-theorem on the end-of-the-world brane $Q$ rather than the holographic c-theorem in the bulk $N$. Thus it is sufficient to focus on an AdS spacetime with a dynamical brane. }
\begin{eqnarray}\label{AdSmetric}
ds^2=\frac{dz^2-dt^2+dx^2+\sum_{a=1}^{d-2}(dy^a)^2 }{z^2}.
\end{eqnarray}
Since there are matter fields on the brane generally, the embedding function (\ref{GBQ}) of the brane and the NBC (\ref{NBCGBbar}) should be replaced by
\begin{eqnarray}\label{branematter}
x=-F(z),
\end{eqnarray}
and 
\begin{eqnarray}\label{NBCmatter}
 \left(1+2\alpha \left(d-1\right)\left(d-2\right)\right)\left(K^{ij}-\left(K-T\right)h^{ij}\right) +2\alpha l^2(Q^{ij}-\frac{1}{3}Q h^{ij})=\frac{1}{2}T^{ij}_{\text{matter}},
\end{eqnarray}
respectively, where $T^{ij}_{\text{matter}}$ is the matter stress tensor.  Imposing the null energy condition on the brane,
\begin{eqnarray}\label{NEN}
T^{ij}_{\text{matter}} N_i N_j\ge 0,
\end{eqnarray}
where
\begin{eqnarray}\label{NtNx}
N^t=\pm 1, \ N^z=\frac{1}{\sqrt{1+(F'(z))^2}},\ N^x=\frac{F'(z)}{\sqrt{1+(F'(z))^2}},\ N^a=0,
\end{eqnarray}
we obtain 
\begin{eqnarray}\label{NEN1}
-  \frac{\left(1+4\alpha(d-2) +(1+4\alpha (d-2)^2) F'(z)^2\right)}{z \left(1+F'(z)^2\right)^{5/2}}F''(z) \ge 0.
\end{eqnarray}
By using the causality constraint (\ref{GBconstraintbar}) with $d\ge 4$, we get
\begin{eqnarray}\label{NEN2}
F''(z) \le 0.
\end{eqnarray}

Now let us discuss the construction of g-function. The g-function should satisfy the following conditions. First, it  decreases under RG flows
\begin{eqnarray}\label{gcondition1}
g'(z) \le 0,
\end{eqnarray}
where $1/z$ denotes the energy scale. Second, on the AdS boundary $z=0$ (UV fixed point), the g-function reduces to the quantity which represents the physical degrees of freedom on the boundary
\begin{eqnarray}\label{gcondition2}
\lim_{z\to 0}g(z) = c_{\text{bdy}},\ S_{\text{bdy}}.
\end{eqnarray}
It can be either the A-type boundary central charge $c_{\text{bdy}}$ or the boundary entropy $S_{\text{bdy}}$.
However, it cannot be the B-type boundary central charge such as (\ref{alphadGB}), which disobeys the g-theorem generally.

\begin{figure}[t]
\centering
\includegraphics[width=10cm]{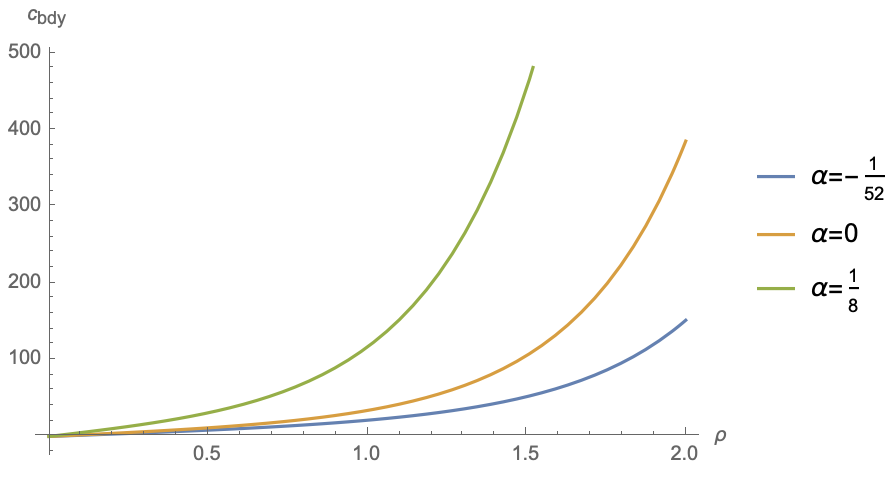}
\includegraphics[width=10cm]{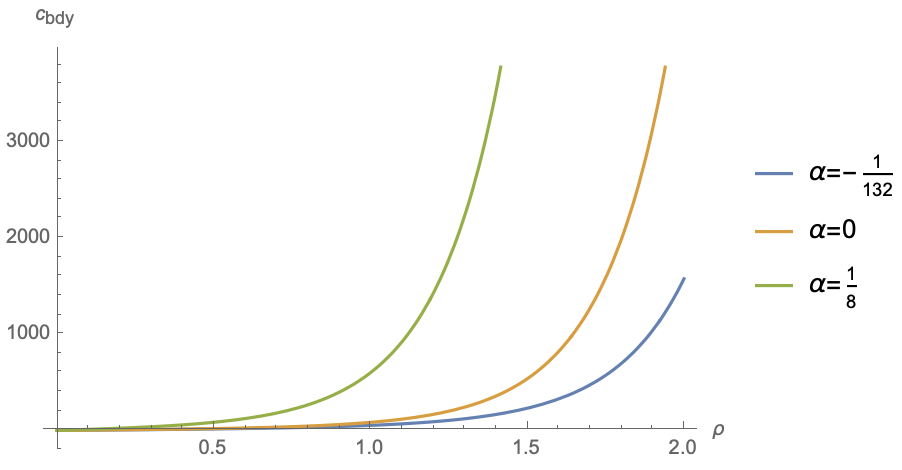}
\caption{A-type boundary central charge $c_{\text{bdy}}$ for $d=5$ (above) and $d=7$ (below). The blue line, orange line and green line correspond to the lower bound, zero and the upper bound of the Gauss-Bonnet coupling $\a$. A-type boundary central charge $c_{\text{bdy}}$ is positive for $\rho>0$ and increases with $\rho$. The larger $\alpha$ is, the larger the A-type boundary central charge is.}
\label{cbdy5d7d}
\end{figure}

We construct the following g-function, which obeys the above two requirements. Multiplying the null energy condition (\ref{NEN1}) by a  positive function 
$S(d-2)\left(1+F'(z)^2\right)^{\frac{d}{2}}$ 
and integrating along $z$, we get
\begin{eqnarray}\label{g2function2}
&&g(z)= S(d-2) \int dz F''(z) \left(1+F'(z)^2\right)^{\frac{d-5}{2}} \left(1+4 \alpha  (d-2)+\left(4 \alpha  (d-2)^2+1\right) F'(z)^2\right) \nonumber\\
&&=\frac{S(d-2)}{d-2}F'(z) \left((d-3) \, _2F_1\left(\frac{1}{2},\frac{5-d}{2};\frac{3}{2};-F'(z)^2\right)+\left(4 \alpha  (d-2)^2+1\right) \left(F'(z)^2+1\right)^{\frac{d-3}{2}}\right),\nonumber\\
\end{eqnarray}
where $S(d-2) =2 \pi ^{\frac{d-1}{2}}/\Gamma \left(\frac{d-1}{2}\right)$ is the volume of a $(d-2)$-dimensional unit sphere.
By construction, $g(z)$ obeys the g-theorem (\ref{gcondition1}). 
 Consider the limit $z\to 0$, we get 
\begin{eqnarray}\label{app:cbdy123}
&&\lim_{z\to 0}g(z)=c_{\text{bdy}}\nonumber\\
&&=\frac{S(d-2)\sinh (\rho )}{d-2} \left(\left(4 \alpha  (d-2)^2+1\right) \cosh ^{d-3}(\rho )+(d-3) \, _2F_1\left(\frac{1}{2},\frac{5-d}{2};\frac{3}{2};-\sinh ^2(\rho )\right)\right),\nonumber\\
\end{eqnarray}
which is just the A-type boundary central charge (\ref{app:cbdy}) defined by Weyl anomaly. Note that the A-type boundary central charge defined by Weyl anomaly appears only for odd $d$. Please see the appendix for the derivations of A-type boundary central charges from holographic Weyl anomaly.  For $d=5$ and $d=7$, we have
\begin{eqnarray}
  \label{g25d7d}
c_{\text{bdy}}=\lim_{z\to 0}g(z)=
\begin{cases}
\frac{2}{3} \pi ^2 \sinh (\rho ) \left((36 \alpha +1) \cosh ^2(\rho )+2\right),& d=5,\\
\frac{1}{15} \pi ^3 \sinh (\rho ) \left((300 \alpha +3) \cosh ^4(\rho )+2 (\cosh (2 \rho )+5)\right),& d=7.
\end{cases}
\end{eqnarray}
Remarkably, as shown in Fig.\ref{cbdy5d7d}, $c_{\text{bdy}}$ increases with $\rho$.  From $F''(z)\ge 0$ (\ref{NEN2}) and $\lim_{z\to0} F'(z)=\sinh(\rho)$, we have $\rho(\text{UV}) \ge \rho(\text{IR}) $.  As a result, we have $c_{\text{bdy}}(\text{UV})\ge c_{\text{bdy}}(\text{IR})$, which is consistent with the g-theorem (\ref{gcondition1}). 

Remarkably, as we will show below, the g-function (\ref{g2function2}) can also reproduce the universal term of boundary entropy of a half ball in general dimensions.  Let us first review the boundary entropy in $\text{AdS}_3/\text{BCFT}_2$ \cite{Takayanagi:2011zk,Fujita:2011fp} and then generalize it to higher dimensions.  In the case of $\text{AdS}_3/\text{BCFT}_2$, \cite{Takayanagi:2011zk,Fujita:2011fp} propose to consider the holographic entanglement entropy of the strip 
\begin{eqnarray}\label{sect2.2:strip}
0\le x \le L,
\end{eqnarray}
where $x=0$ denoted the boundary of BCFT, and then obtain the boundary entropy as
\begin{eqnarray}\label{sect2.2:2dSbdy}
S_{\text{bdy}}=S_{\text{HEE}}(\rho)-S_{\text{HEE}}(0)=\frac{\rho}{4G_N}.
\end{eqnarray}
In higher dimensions, naturally one may still consider the holographic entanglement entropy of the strip $0\le x \le L$ and then derive boundary entropy by $S_{bdy}=S_{\text{HEE}}(\rho)-S_{\text{HEE}}(0)$.  However, unlike the case of $d=2$, in higher dimensions, the RT surface of the strip $0\le x \le L$ cannot end on the end-of-the-world brane for large enough tension $\rho>\rho_c$ \cite{Geng:2021mic,Chu:2017aab}, 
which yields a ill-defined boundary entropy. In fact, in higher dimensions, \cite{Fujita:2011fp} proposes to use $\rho$ (or equivalently, the brane tension) instead of the boundary entropy of a strip as the g-function.  

Inspired by wedge holography \cite{Akal:2020wfl}, we propose to consider the boundary entropy of a half ball 
\begin{eqnarray}\label{halfball}
x^2+y_a^2\le L^2, \ x\ge 0,
\end{eqnarray}
instead of the strip $0\le x \le L$.  This proposal has many advantages. First, it can resolve the problem found in \cite{Geng:2021mic,Chu:2017aab} so that the RT surface always ends on the brane for general brane tensions.  Second, the half ball (\ref{halfball}) reduces to the strip $0\le x \le L$ for $d=2$ (since $y_a=0$ for $d=2$. see eq.(\ref{AdSmetric})). Thus it includes the 2d case of \cite{Takayanagi:2011zk,Fujita:2011fp} as a special case.  
Recall that we have $y_a$=0 for $d=2$. 
Third, as we will show below, the boundary entropy of a half ball exactly agrees with the g-function and A-type boundary central charge. 

Performing the coordinate transformations $z=Z /\cosh(r), \ x=Z \tanh(r),\ y_a^2=R^2$, the AdS metric (\ref{AdSmetric}) becomes
\begin{eqnarray}\label{AdSmetricbrane}
 ds^2=dr^2+\cosh^2 (r) \frac{dZ^2-dt^2+ dR^2+ R^2 d\Omega^2}{Z^2}, \ \  -\rho\le r<\infty
\end{eqnarray}
where the brane is located at $r=-\rho$ and the AdS boundary is at $r=\infty$. Now the half ball (\ref{halfball}) on the AdS boundary becomes
\begin{eqnarray}\label{halfball1}
Z^2+ R^2 \le L^2. 
\end{eqnarray}
According to \cite{Akal:2020wfl}, the RT surface for the half ball (\ref{halfball}) is given by 
\begin{eqnarray}\label{sec2.2:RTsurface}
Z^2+R^2=L^2, \ \ t=\text{constant},
\end{eqnarray}
for Einstein gravity. One can check that it still works for Gauss-Bonnet gravity (due to the fact that all of the components of the extrinsic curvatures of the RT surface (\ref{sec2.2:RTsurface}) vanish).  From the action of Gauss-Bonnet gravity (\ref{IGBmiao}) and setting $16\pi \bar{G}_N=l=1$, we obtain the holographic entanglement entropy 
\begin{eqnarray}\label{HEE}
S_{\text{HEE}}&=&4\pi \int_{\Gamma} d^{d-1}x \sqrt{\gamma} \Big((1+2\alpha(d-1)(d-2))+2 \alpha R_{\gamma} \Big)\nonumber\\
&+& 4\pi \int_{\partial \Gamma} d^{d-2}x \sqrt{\sigma}\  4\alpha K_{\partial \Gamma},
 \end{eqnarray}
 where $\Gamma$, $\gamma$ and $R_{\gamma}$ denote the RT surface, the induced metric and intrinsic Ricci scalar on the RT surface, respectively. Similarly, $\partial \Gamma$, $\sigma$ and $K_{\partial \Gamma}$  are the boundary of RT surface (the intersection of the RT surface and the brane), the induced metric and extrinsic curvature on the boundary of RT surface, respectively. 
 From (\ref{AdSmetricbrane},\ref{sec2.2:RTsurface}), we derive for $d>2$ \footnote{Note that there is no coordinate $R$ for $d=2$, thus formulas (\ref{sect2.2:formulas}) works only for $d\ge 3$. For $d=2$, the half ball (\ref{halfball}) reduces to the strip (\ref{sect2.2:strip}) and the boundary entropy is given by (\ref{sect2.2:2dSbdy}). }
 \begin{eqnarray}\label{sect2.2:formulas}
 && R_{\gamma}=-(d-1)(d-2),\ \ \  K_{\partial \Gamma}= (d-2) \tanh (\rho ),\nonumber\\
&& \int_{\Gamma} dx^{d-1} \sqrt{\gamma}=\int_{-\rho}^{\infty} \cosh ^{d-2}(r) dr  \int L R^{d-3} \left(L^2-R^2\right)^{\frac{1}{2}-\frac{d}{2}} dR d\Omega, \\
&& \int_{\partial \Gamma} dx^{d-2} \sqrt{\sigma}= \cosh ^{d-2}(\rho) \int L R^{d-3} \left(L^2-R^2\right)^{\frac{1}{2}-\frac{d}{2}} dR d\Omega.\nonumber
 \end{eqnarray}
 Substituting the above formulas into (\ref{sect2.2:2dSbdy}, \ref{HEE}), we derive the boundary entropy for $d> 2$
 \begin{eqnarray}\label{sect2.2:bdyentropy1}
&&S_{\text{bdy}}(\rho)=S_{\text{HEE}}(\rho)-S_{\text{HEE}}(0)\nonumber\\
&&= \Big(16 \pi  \alpha  (d-2) \sinh (\rho ) \cosh ^{d-3}(\rho ) + 4 \pi \int_{-\rho}^0 \cosh ^{d-2}(r) dr\Big) \int L R^{d-3} \left(L^2-R^2\right)^{\frac{1}{2}-\frac{d}{2}} dR d\Omega. \nonumber\\
\end{eqnarray}
We notice that the boundary entropy (\ref{sect2.2:bdyentropy1}) is just one half \footnote{The factor half is due to the fact that we have $-\rho\le r\le 0$ for boundary entropy, while $-\rho\le r\le \rho$ for wedge holography \cite{Akal:2020wfl}.} of the holographic entanglement entropy of 
a disk in wedge holography. See  eq.(2.40) of  \cite{Akal:2020wfl}. And the prefactor of (\ref{sect2.2:bdyentropy1}) is just the inverse of the effective Newton's constant  
$\frac{1}{4G_{\text{eff}}}$
 on the brane \cite{Akal:2020wfl,Miao:2020oey}. According to double holography, the boundary state of BCFT is dual to the gravity on the brane. Thus it is natural that the entanglement entropy of boundary state of BCFT (boundary entropy) is 
inversely proportional to the effective Newton's constant of gravity on the brane.

We are interested in the universal terms of boundary entanglement entropy (\ref{sect2.2:bdyentropy1}), which are given by the UV 
logarithmic divergent terms for odd $d$, and are given by finite terms for even $d$ \cite{Ryu:2006ef,Myers:2010tj}
 \begin{eqnarray}\label{sect2.2:universalEE}
S_{\text{bdy\ univ}}=
\begin{cases}
(-1)^{\frac{d-1}{2}-1} 4 \ c_{\text{bdy}} \log(\frac{2 L}{\epsilon}),&\ \text{for odd } d,\\
(-1)^{\frac{d-2}{2}} 2\pi \  c_{\text{bdy}} ,& \ \text{for even }d,
\end{cases}
\end{eqnarray}
where $\epsilon$ is the UV cut-off, 
$L$ is the radius of the half-ball and $c_{\text{bdy}}$ is the A-type boundary central charge.   Recall that the A-type boundary central charge $c_{\text{bdy}}$ defined by Weyl anomaly (\ref{Weylanomaly}) applies to only odd 
$d$. On the other hand, $c_{\text{bdy}}$ defined by the universal term of boundary entropy (\ref{sect2.2:universalEE}) works well in general dimensions. As we will show below, these two definitions are consistent. 

Performing the coordinate transformation $R=L \sqrt{y^2-1}/y$ and focusing on the universal terms, we obtain
 \begin{eqnarray}\label{sect2.2:universalformula}
&& \int L R^{d-3} \left(L^2-R^2\right)^{\frac{1}{2}-\frac{d}{2}} dRd\Omega|_{\text{univ}}=S(d-3) \int (y^2-1)^{\frac{d-4}{2}} dy|_{\text{univ}}\nonumber\\
&=& \frac{S(d-2)}{4\pi} \begin{cases}
(-1)^{\frac{d-1}{2}-1} 4\log(\frac{2 L}{\epsilon}),&\ \text{for odd } d,\\
(-1)^{\frac{d-2}{2}} 2\pi ,& \ \text{for even }d,
\end{cases}
\end{eqnarray}
where $S(d-2) $ is the volume of a $(d-2)$-dimensional unit sphere. In the above derivations, we have used eq.(2.10) and eq.(2.11) of \cite{Hung:2011nu}. 

From (\ref{sect2.2:bdyentropy1},\ref{sect2.2:universalEE},\ref{sect2.2:universalformula}), we finally derive the boundary central charge defined by the universal term of boundary entropy
 \begin{eqnarray}\label{sect2.2:bdyentropy2}
c_{\text{bdy}}=\frac{S(d-2)}{4\pi}\Big(16 \pi  \alpha  (d-2) \sinh (\rho ) \cosh ^{d-3}(\rho ) + 4 \pi \int_{-\rho}^0 \cosh ^{d-2}(r) dr\Big).
\end{eqnarray}
By using the formulas
 \begin{eqnarray}\label{sect2.2:intrgral}
\int \cosh^{d-2}(r) dr&=&\int (1+x^2)^{\frac{d-3}{2}} dx=x \, _2F_1\left(\frac{1}{2},\frac{3-d}{2};\frac{3}{2};-x^2\right)\nonumber\\
&=&\frac{x \left((d-3) \, _2F_1\left(\frac{1}{2},\frac{5-d}{2};\frac{3}{2};-x^2\right)+\left(x^2+1\right)^{\frac{d-3}{2}}\right)}{d-2},
\end{eqnarray}
where $x=\sinh(r)$, we simplify the boundary central charge (\ref{sect2.2:bdyentropy2}) as
 \begin{eqnarray}\label{sect2.2:bdyentropy3}
&&c_{\text{bdy}}=\lim_{z\to 0}g_2(z)\nonumber\\
&&=\frac{S(d-2)\sinh (\rho )}{d-2} \left(\left(4 \alpha  (d-2)^2+1\right) \cosh ^{d-3}(\rho )+(d-3) \, _2F_1\left(\frac{1}{2},\frac{5-d}{2};\frac{3}{2};-\sinh ^2(\rho )\right)\right),\nonumber\\
\end{eqnarray}
which exactly agrees with the g-function (\ref{g2function2},\ref{app:cbdy123}) in general dimensions and A-type boundary central charge (\ref{app:cbdy}) defined by Weyl anomaly for odd $d$. 

Now we finish the proof of holographic g-theorem for Gauss-Bonnet gravity.  And we have verified that our g-function can produce 
the universal term 
of boundary entropy of a half ball in general dimensions and the A-type boundary central charges defined by Weyl anomaly for odd $d$.  These are strong support for our results.

\section{AdS/BCFT for curvature-squared gravity}

In this section, we formulate AdS/BCFT for curvature-squared gravity. 
We first review the overly restrictive problem for curvature-squared gravity in sect.3.1 and then give two resolutions to this problem in sect.3.2 and sect.3.3, respectively. In particular, we show that the holographic g-theorem is obeyed by general curvature-squared gravity.

\subsection{The problem}

Let us start with the first-order action of general higher curvature gravity \cite{Deruelle:2009zk}
\begin{eqnarray}\label{fRaction}
I&=&\int_{N}d^{d+1}x\sqrt{|g|} \Big[  f(\phi_{\mu\nu\rho\sigma})+\Psi^{\mu\nu\rho\sigma} (R_{\mu\nu\rho\sigma}-\phi_{\mu\nu\rho\sigma})\Big] \nonumber\\
&+& \int_{Q}d^{d}y\sqrt{|h|}( 4\Psi_{\mu\nu} K^{\mu\nu}-2 T),
\end{eqnarray}
where $K_{\mu\nu}=h^{\alpha}_{\mu}h^{\beta}_{\nu}\nabla_{\alpha}n_{\beta}$ is the extrinsic curvature, $T$ is the brane tension, $n_{\mu}$ is the outward-directed normal vector, $h_{\mu\nu}=g_{\mu\nu}-n_{\mu}n_{\nu}$ is the induced metric on the boundary and 
\begin{eqnarray}\label{fRPsi}
\Psi_{\mu\nu}=\Psi_{\mu\rho\nu\sigma} n^{\rho} n^{\sigma},\ \ \Psi^{\mu\nu\rho\sigma}=\frac{\partial f}{\partial R_{\mu\nu\rho\sigma}}.
\end{eqnarray}
Taking the variation of the action and focusing on the boundary terms, we have 
\begin{eqnarray}\label{fRdI}
&&\delta I= 4\int_{Q}d^{d}y\sqrt{|h|} K^{\mu\nu} \delta \Psi_{\mu\nu}\nonumber\\
&&+ \int_{Q}d^{d}y\sqrt{|h|} \Big( 2 n^{\alpha} \nabla^{\beta} \Psi_{\mu\beta\alpha\nu}+6 \Psi_{\alpha\mu} K^{\alpha}_{\nu} +2 D^{\alpha} \Psi^{\parallel}_{\nu\alpha\mu} -h_{\mu\nu}( 2\Psi_{\alpha\beta} K^{\alpha\beta}- T) \Big)^{\parallel}\delta h^{\mu\nu}
\end{eqnarray}
where $D_{\alpha}$ is the covariant derivative defined by $h_{\mu\nu}$, $\Psi^{\parallel}_{\nu\alpha\mu}=n^{\beta} \Psi_{\nu_1\alpha_1\beta\mu_1}h^{\nu_1}_{\nu}h^{\alpha_1}_{\alpha}h^{\mu_1}_{\mu}$ and $\parallel$ denotes the tangential direction.  See \cite{Jiang:2018sqj} for the derivations. 

Imposing NBC for both $h^{\mu\nu}$ and $\Psi_{\mu\nu}$, we get 
\begin{eqnarray}\label{fRNBC1}
&& K^{\mu\nu}=0,\\ \label{fRNBC2}
&& \Big( 2 n^{\alpha} \nabla^{\beta} \Psi_{\beta(\mu\nu)\alpha}+6 \Psi_{\alpha(\mu} K^{\alpha}_{\nu)} -2 D^{\alpha} \Psi^{\parallel}_{\alpha(\nu\mu)} -h_{\mu\nu}( 2\Psi_{\alpha\beta} K^{\alpha\beta}- T) \Big)^{\parallel}=0.
\end{eqnarray}
As we have mentioned in the introduction, $K^{\mu\nu}=0$ is too constrained. For the AdS metric (\ref{AdSmetric}) and the embedding function of the brane $x=-\sinh(\rho) z$, we have $K_{\mu\nu}=\tanh(\rho) h_{\mu\nu}$.  As a result, $K_{\mu\nu}=0$ leads to $\rho=0$, which yields zero boundary entropy \cite{Takayanagi:2011zk} and the A-type boundary central charges (\ref{app:cbdy123}). This means that there are no boundary degrees of freedom and the dual BCFT is trivial.  Besides, (\ref{fRNBC2}) does not agree with the NBC (\ref{NBCGB},\ref{NBCGBbar}) of Gauss-Bonnet gravity.

\subsection{Resolution I}

We aim to resolve the above problems. For simplicity, we focus on curvature-squared gravity
 \begin{eqnarray}\label{Ibulkcurvature}
I_{\text{bulk}}=\int_N d^{d+1}x\sqrt{|g|} \Big(R+d(d-1)+\alpha\ \mathcal{L}_{\text{GB}}(\bar{R}) +\bar{c}_1 \bar{R}^2+\bar{c}_2 \bar{R}_{\mu\nu}\bar{R}^{\mu\nu}\Big),
 \end{eqnarray}
 where the background curvature $\bar{R}$ is defined by (\ref{background curvature1},\ref{background curvature2},\ref{background curvature3}). Note that the causality constraints are still given by (\ref{GBconstraintbar}) for general curvature-squared gravity \cite{Sen:2014nfa}. 
In order to be consistent with the case of Gauss-Bonnet gravity, we propose to add the standard boundary term  (\ref{IGBmiao}) for $R$ and $\mathcal{L}_{\text{GB}}(\bar{R})$, while add the GHY-like term (\ref{fRaction}) for $\bar{R}^2$ and $\bar{R}_{\mu\nu}\bar{R}^{\mu\nu}$  \begin{eqnarray}\label{Ibdycurvature}
I_{\text{bdy}}=2\int_Q d^{d}y\sqrt{|h|} \Big( (1+2\alpha(d-1)(d-2))(K-T)+ 2\alpha (J-2 G^{ij}_{Q} K_{ij}) + 2 \Psi_{ij} K^{ij}\Big),
 \end{eqnarray}
 where 
\begin{eqnarray}\label{fRPsiij1}
&&\Psi_{ij}=\Psi_{\mu\rho\nu\sigma} n^{\rho} n^{\sigma}\frac{\partial x^{\mu}}{\partial y^i}\frac{\partial x^{\nu}}{\partial y^j},\\ \label{fRPsiij2}
&&\Psi_{ijk}=\Psi_{\mu\nu\rho\sigma} n^{\rho} \frac{\partial x^{\mu}}{\partial y^i}\frac{\partial x^{\nu}}{\partial y^j}\frac{\partial x^{\sigma}}{\partial y^k},\\ \label{fRPsiij3}
&&\Psi_{\mu\nu\rho\sigma}=\bar{c}_1 \bar{R}(g_{\mu\rho}g_{\nu\sigma}-g_{\mu\sigma}g_{\nu\rho})+\frac{\bar{c}_2}{2}(\bar{R}_{\mu\rho}g_{\nu\sigma}-\bar{R}_{\mu\sigma}g_{\nu\rho}+g_{\mu\rho}\bar{R}_{\nu\sigma}-g_{\mu\sigma}\bar{R}_{\nu\rho}).
\end{eqnarray}
Taking variations of the total action $I=I_{\text{bulk}}+I_{\text{bdy}}$ and focusing on the boundary terms, we obtain 
\begin{eqnarray}\label{res:fRdI}
\delta I= \int_{Q}d^{d}y\sqrt{|h|}( 4K^{ij} \delta \Psi_{ij}-\frac{1}{2} T^{\text{HD}}_{ij}\delta h^{ij}),
\end{eqnarray}
where 
\begin{eqnarray}\label{res:Tij}
-\frac{1}{2}T^{\text{HD}}_{ij}&=&\left(1+2\alpha \left(d-1\right)\left(d-2\right)\right)\left(K_{ij}-\left(K-T\right)h_{ij}\right) +2\alpha(Q_{ij}-\frac{1}{3}Q h_{ij})\nonumber\\
&&+ 2 h^{\mu}_i h^{\nu}_j n^{\alpha} \nabla^{\beta} \Psi_{\beta(\mu\nu)\alpha}+6 \Psi_{l(i} K^{l}_{j)} -2 D^{k} \Psi_{k(ij)} -2h_{ij}\Psi_{kl} K^{kl},
\end{eqnarray}
and $h^{\mu}_i =\partial x^{\mu}/\partial y^i$ is the projection operator. 
To relax the constraint on the shape of the brane, we propose to impose DBC for the auxiliary field $\Psi_{ij}$
\begin{eqnarray}\label{res:DBC}
\delta \Psi_{ij}|_Q=0,
\end{eqnarray}
and impose NBC for the induced metric $h_{ij}$
\begin{eqnarray}\label{res:NBC}
T^{\text{HD}}_{ij}|_Q=0.
\end{eqnarray}
Let us make some comments on the DBC (\ref{res:DBC}). 
First, naturally, we require that AdS is a solution to AdS/BCFT for higher curvature gravity. Since $\bar{R}=\bar{R}_{\mu\nu}=0$ for AdS, from (\ref{fRPsiij1},\ref{fRPsiij2},\ref{fRPsiij3}) we get
\begin{eqnarray}\label{res:DBCforAdS}
 \Psi_{ij}|_Q=0.
\end{eqnarray}
Second, different choices of BCs yield different datas for the dual BCFTs, such as one point functions, boundary central charges and so on \cite{Miao:2018qkc,Chu:2021mvq}. In general, quantities on the brane contribution to one point functions for DBC  \cite{Miao:2018qkc}. Third, as we will show below, DBC (\ref{res:DBC}) impose non-trivial BC only for the massive graviton. In this paper, we mainly focus on the massless gravitational modes. We leave the careful study of the one point function of massive modes related to DBC (\ref{res:DBC}) in future works.

 In the transverse traceless gauge \footnote{ We are not interested in the scalar mode $g^{\mu\nu}\delta g_{\mu\nu}$ in higher curvature gravity. Instead, we focus on the gravitational modes. }
\begin{eqnarray}\label{res:gauge}
g^{\mu\nu}\delta g_{\mu\nu}=0, \  \nabla^{\mu} \delta g_{\mu\nu}=0,
\end{eqnarray}
the linear perturbation equation of curvature-squared gravity (\ref{Ibulkcurvature}) is given by
\begin{eqnarray}\label{res:EOMRR}
(\Box+2)\Big(\bar{c}_2(\Box+2)+(1+4 (d-2)\a)\Big)\delta g_{\mu\nu}=0,
\end{eqnarray}
where $\Box$ is the D'Alembert operator defined by the AdS metric with radius $l=1$. From (\ref{res:EOMRR}), we observe that there is a massless graviton obeying 
\begin{eqnarray}\label{res:masslessmode}
(\Box+2)\delta g_{\mu\nu}=0,
\end{eqnarray}
and a massive graviton obeying 
\begin{eqnarray}\label{res:massivemode}
(\Box+2-m^2)\delta g_{\mu\nu}=0,
\end{eqnarray}
 where
\begin{eqnarray}\label{res:mass}
m^2=-\frac{1+4 (d-2)\alpha}{\bar{c}_2}
\end{eqnarray}
is the mass squared.  In order to satisfy the Breitenlohner-Freedman bound $m^2\ge -d^2/4$, we require that
\begin{eqnarray}\label{res:tachyon}
\bar{c}_2<0, \ \ \text{or}\ \ \bar{c}_2 \ge \frac{4(1+4 (d-2)\alpha)}{d^2}. 
\end{eqnarray}
It is well-known that there is ghost in curvature-squared gravity \cite{Deruelle:2009zk}, since the massless mode and the massive mode cannot both have the correct sign of kinetic energy. For our case, the massive graviton is a ghost.  

We are interested in the unitary massless mode (\ref{res:masslessmode}), which is related to the one point function of stress tensor and the holographic g-theorem.  Note that the operator dual to the massive graviton has conformal dimension $\Delta=\frac{d}{2}+\sqrt{\frac{d^{2}}{4}+m^{2}}$, which is irrelevant to the stress tensor with conformal dimension $d$.  Note also that, to prove the holographic g-theorem,  we only need an AdS background (\ref{AdSmetric}) together with a dynamical brane (\ref{branematter}). Thus the gravitational fluctuations in the bulk are irrelevant to the holographic g-theorem on the end-of-the-world brane. 

Considering the first-order perturbation around an AdS background with the gauge (\ref{res:gauge}), from (\ref{fRPsiij1},\ref{fRPsiij3}) we derive
\begin{eqnarray}\label{res:DBC1}
\delta \Psi_{ij}=-\frac{\bar{c}_2}{4} (h^{\mu}_i h^{\nu}_j+h_{ij} n^{\mu}n^{\nu}) (\Box+2) \delta g_{\mu\nu},
\end{eqnarray}
which vanishes for the massless mode (\ref{res:masslessmode}) automatically.  Thus, the DBC $\delta \Psi_{ij}|_Q=0$ (\ref{res:DBC}) imposes boundary condition only on the massive mode.  Note that the NBC (\ref{res:NBC}) for the induced metric does not impose further conditions for the massive mode. Consider the metric and the embedding function of $Q$
\begin{eqnarray}\label{res:metricexample}
&& ds^2=dr^2+\cosh^2 (r) \left( \bar{h}^{(0)}_{ij}(y) + \epsilon H(r) \bar{h}^{(1)}_{ij}(y)  \right)dy^i dy^j+O(\epsilon^2),\\
&& Q: r=-\rho +O(\epsilon^2) \label{res:Qexample}
\end{eqnarray}
where $\bar{h}^{(0)}_{ij}(y)$ is an AdS metric and $\bar{h}^{(1)}_{ij}(y)$ denotes the perturbation. One can check that DBC (\ref{res:DBC}) for the  auxiliary field and NBC (\ref{res:NBC}) for the induced metric yield the same boundary condition for the massive mode
\begin{eqnarray}\label{res:BCmassivemode}
H(-\rho)=0.
\end{eqnarray}
The NBC (\ref{res:NBC}) for the induced metric imposes non-trivial boundary conditions only for the massless mode. In this sense, we say that we impose DBC  (\ref{res:DBC}) for the massive graviton, while imposing NBC (\ref{res:NBC}) for the massless graviton. Now let us focus on the NBC (\ref{res:NBC}) for the massless mode. Since $\Psi_{\mu\nu\rho\sigma}=\bar{R}_{\mu\nu}=\bar{R}=O(\e^2)$, the NBC (\ref{res:NBC}) of curvature-squared gravity is exactly the same as that of Gauss-Bonnet gravity (\ref{NBCGBbar}) at the linear perturbation.  In other words, at the first-order perturbation, curvature-squared gravity and Gauss-Bonnet gravity have the same EOM $(\Box+2)\delta g_{\mu\nu}=0$ and the same boundary condition  (\ref{NBCGBbar}).  As a result, the A-type boundary central charge (\ref{alphadGB}), the one point function of stress tensor (\ref{TijGB1}) and the holographic g-function (\ref{g2function2}) are exactly the same for curvature-squared gravity and Gauss-Bonnet gravity \footnote{They are the same in parameters $(\bar{G}_N, l, \bar{c}_i,\alpha)$. If we take the traditional parameters $(G_N, L, c_i, \lambda_{\text{GB}})$ instead, the central charges and g-functions are different up to some reparameterization. That is because the relations between $(\bar{G}_N, l, \bar{c}_i,\alpha)$ and $(G_N, L, c_i, \lambda_{\text{GB}})$ are different for curvature-squared gravity with $c_i, \bar{c}_i\ne 0$ and Gauss-Bonnet gravity with $c_i=\bar{c}_i=0$. }.  In particular, the holographic g-theorem is obeyed by general curvature-squared gravity, provided that the boundary conditions (\ref{res:DBC},\ref{res:NBC}) are imposed. 

\subsection{Resolution II}

In this section, we provided an alternative resolution to the overly restrictive problem. We find that, by adding suitable boundary terms, it is possible to impose NBCs for both the auxiliary field and the induced metric. The price is that the boundary term does not reduce to that of Gauss-Bonnet gravity when the gravitational couplings are chosen to be those of Gauss-Bonnet gravity. For simplicity, we focus on $\text{AdS}_5/\text{BCFT}_4$ in this subsection. 

For curvature-squared gravity (\ref{Ibulkcurvature}), we proposed to choose the following boundary terms
\begin{eqnarray}\label{resII:Ibdycurvature}
I_{\text{bdy}}=2\int_Q d^{4}y\sqrt{|h|} \Big( K-T+ 2 \Psi_{ij} (K^{ij}+ b_1\Psi^{ij} +b_2 h^{ij} )\Big),
 \end{eqnarray}
where $b_1$ and $b_2$ are constants to be determined, and $\Psi_{ij}$ is given by (\ref{fRPsiij1}) with
\begin{eqnarray}\label{resII:Psii}
\Psi_{\mu\nu\rho\sigma}&=&2\alpha \bar{R}_{\mu\nu\rho\sigma}+(\bar{c}_1+\a) \bar{R}(g_{\mu\rho}g_{\nu\sigma}-g_{\mu\sigma}g_{\nu\rho})\nonumber\\
&&+\frac{\bar{c}_2-4\a}{2}(\bar{R}_{\mu\rho}g_{\nu\sigma}-\bar{R}_{\mu\sigma}g_{\nu\rho}+g_{\mu\rho}\bar{R}_{\nu\sigma}-g_{\mu\sigma}\bar{R}_{\nu\rho}).
\end{eqnarray}
Recall that, for AdS, $\Psi_{ij} =0$ and (\ref{resII:Ibdycurvature}) reduces to the GHY term of Einstein gravity. 
From the variations of the total action, we read off the NBC for the auxiliary field $\Psi_{ij}$
\begin{eqnarray}\label{resII:NBC1}
K^{ij}+ 2b_1\Psi^{ij} +b_2 h^{ij} =0,
\end{eqnarray}
and NBC for the induced metric $h^{ij}$
\begin{eqnarray}\label{resII:NBC2}
&&K_{ij}-\left(K-T\right)h_{ij} -2h_{ij}\Psi_{kl}( K^{kl}+ b_1\Psi^{kl} +b_2 h^{kl})\nonumber\\
&&+ 2 h^{\mu}_i h^{\nu}_j n^{\alpha} \nabla^{\beta} \Psi_{\beta(\mu\nu)\alpha}+6 \Psi_{l(i} K^{l}_{j)} -2 D^{k} \Psi_{k(ij)} +4 b_2 \Psi_{ij}+8 b_1 \Psi_{in}\Psi^{n}_j=0,
\end{eqnarray}
respectively. 

Let us study the one point function of stress tensor for the second model of curvature-squared gravity (\ref{Ibulkcurvature},\ref{resII:Ibdycurvature}). We follow the approach of sect.2.1. As explained in the above subsection, we only need to consider the massless mode, which obeys the same EOM (\ref{res:masslessmode}) as Einstein gravity and Gauss-Bonnet gravity at the 
first-order of $O(\e)$. Substituting the bulk metric  (\ref{GBmetric}) into EOM (\ref{res:masslessmode}), we get one independent equation (\ref{GBeq}), which can be solved as (\ref{GBsolution}).  So far, everything is the same as that of sect.2.1. This happens because we have used the background-field method  \cite{Miao:2013nfa}.

Now imposing BCs (\ref{resII:NBC1},\ref{resII:NBC2}) on the brane (\ref{GBQ}), we solve 
\begin{eqnarray}\label{resII:tension}
T=3 \tanh (\rho ), \ \ b_2=-\tanh (\rho ),
\end{eqnarray}
at the leading order $O(\e^0)$, and derive
\begin{eqnarray}\label{resII:alpha}
b_1=\tanh (\rho ),\ \ \ \alpha_4=\frac{2 \cosh ^3(\rho )[1+4(d-2)\alpha] }{-16 \alpha  \sinh (\rho )+\sinh (\rho )+\sinh (3 \rho )+3 \cosh (\rho )+\cosh (3 \rho )},
\end{eqnarray}
at the sub-leading order $O(\e)$.  Note that the brane tension $T$ and the Casimir coefficient $\a_4$ are different for model I of sect. 3.2 and model II of this subsection.  This is not surprising because model I and model II have different boundary terms and boundary conditions. For general curvature-squared gravity (\ref{Ibulkcurvature}), the holographic formula of stress tensor is still given by (\ref{holoTijGB}), from which we finally obtain the one point function of stress tensor (\ref{TijGB1}) with the Casimir coefficient given by (\ref{resII:alpha}). 

\begin{figure}[t]
\centering
\includegraphics[width=10cm]{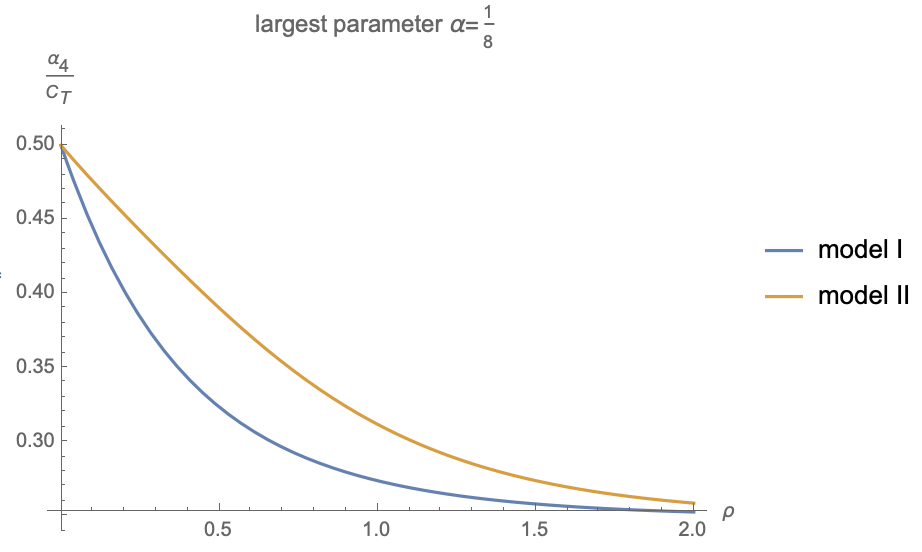}
\includegraphics[width=10cm]{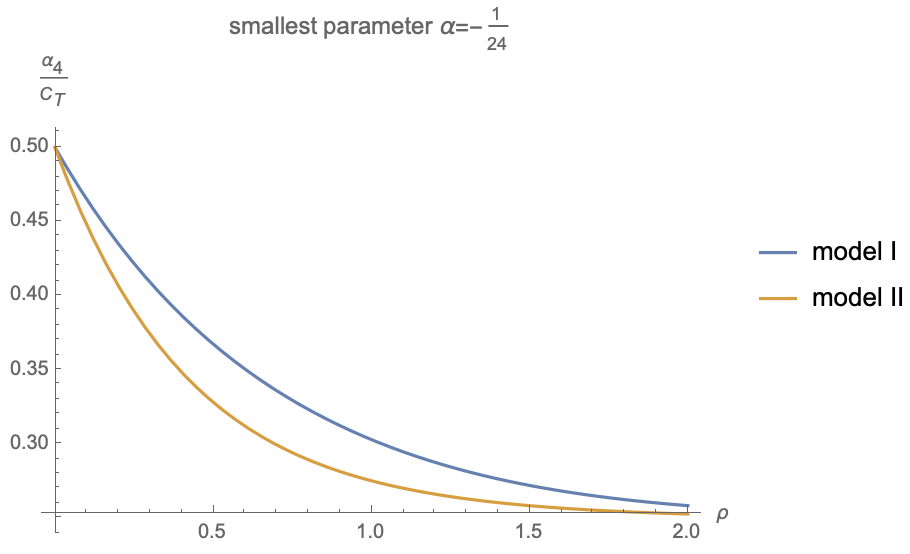}
\caption{Casimir coefficient $\a_4$ for the largest parameter $\a=\frac{1}{8}$ (above) and the smallest parameter $\a=-\frac{1}{24}$ (below). The blue line and orange line correspond to model I of sect.3.2 and model II of sect.3.3, respectively. The Casimir coefficient $\a_d$ is positive and decreases with the brane tension $\rho$. }
\label{adRRmodelII}
\end{figure}

Let us make some comments. {\bf 1}. The NBCs (\ref{resII:NBC1},\ref{resII:NBC2}) can indeed relax the constrain on the shape of the brane and yield non-trivial shape dependence of the stress tensor  (\ref{TijGB1}). However, the price is that we need to add complicated boundary terms, which do not agree with the standard boundary terms of Gauss-Bonnet gravity. {\bf 2}. As shown in Fig.\ref{adRRmodelII}, the B-type central charge $\a_4$ (\ref{resII:alpha}) is positive, which is a strong support for model II.  {\bf 3}.  On an AdS background, the action of the curvature-squared gravity (\ref{Ibulkcurvature},\ref{resII:Ibdycurvature}) of model II is exactly the same as that of Einstein gravity. As a result, the A-type boundary central charge and holographic g-theorem are also the same as those of Einstein gravity. {\bf 4}.  To summarize, the two models of higher curvature gravity in sect.3.2 and sect.3.3 are both well-defined, since they  both have positive B-type central charges and both obey the holographic g-theorem. 

\section{Island and Page curve}

Recently, it is found that the Page curve of Hawking radiation can be recovered due to the emergence of  island in black holes \cite{Penington:2019npb,Almheiri:2019psf,Almheiri:2020cfm}.  In this section, we investigate the version of information paradox for eternal black holes in AdS/BCFT of higher curvature gravity.  Since the exact black hole solutions are only known for Gauss-Bonnet gravity, we focus on Gauss-Bonnet gravity in this section. We verify that,  due to the island, the Page curve can indeed be recovered for the eternal Gauss-Bonnet  black hole. Interestingly, the Page-curve is discontinuous for one range of couplings obeying the causality constraints, which implies that there are 
zeroth-order phase transitions of entanglement entropy. 

Let us first briefly review the version of information paradox for eternal black holes \cite{Almheiri:2019yqk,Almheiri:2019psy}.  The eternal two-sided black hole is dual to  the thermofield double state of CFTs \cite{Eternal black hole M}
\begin{eqnarray}\label{TFD}
| \text{TFD}\rangle=Z^{-1/2} \sum_{\alpha} e^{-E_{\a}/(2T)}e^{-iE_{\a}(t_L+t_R)} | E_{\a}\rangle_L  | E_{\a}\rangle_R,
\end{eqnarray}
where $L$ and $R$ label the states (times) associated with the left and right boundaries.  Couple the eternal two-side black hole to a bath on each side. The system keeps invariant if we move time forward on one side 
while moving time backward on the other side so that $t_L+t_R=0$. On the other hand, the system evolutes if we move time forward on both sides.  What is the time evolution of entanglement entropy of the union of two baths? Without the island, the entanglement entropy will increase forever.  That is because the Hawking radiation enter the bath and their entangled partners fall into the black hole. As a result, the entanglement between black holes and baths keep increasing. At late times, the entanglement entropy will exceed the double black hole entropy. However, the fine-grained entanglement entropy cannot be larger than the coarse-grained black hole entropy \cite{Almheiri:2020cfm}.   This is the version of information paradox for eternal black holes.

\begin{figure}[t]\centering
\includegraphics[width=10cm]{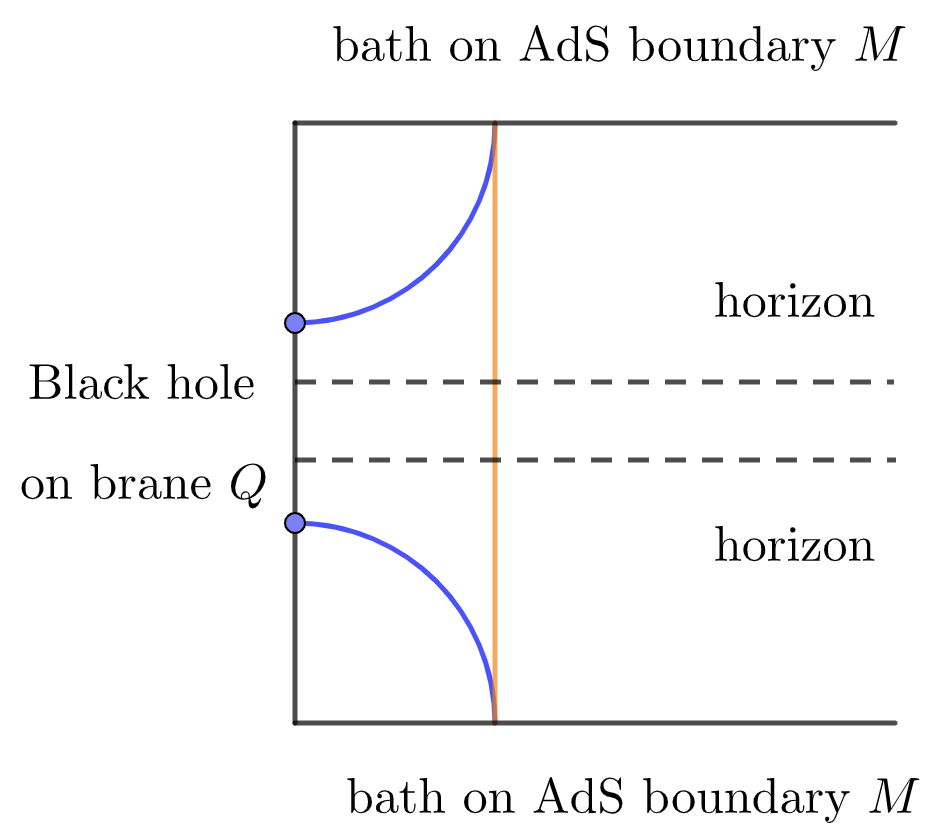}
\caption{Doubly holographic setup (space-like slice): Black hole lives on the end-of-the-world brane $Q$ ($x=0$), and bath lives on the AdS boundary $M$ ($z=0$). The 
dotted line denotes the horizon of black hole, the blue point labels the location of island outside horizon. There are two kinds of extremal surfaces in the bulk. The orange extremal surface passing through the horizon depends on time and is dominant at early times. See Fig.\ref{penrose}. The blue extremal surface ending on the brane is 
time-independent and is dominant at late times. It is the 
so-called island.}
\label{doubleholo}
\end{figure}

\begin{figure}[t]\centering
\includegraphics[width=9cm]{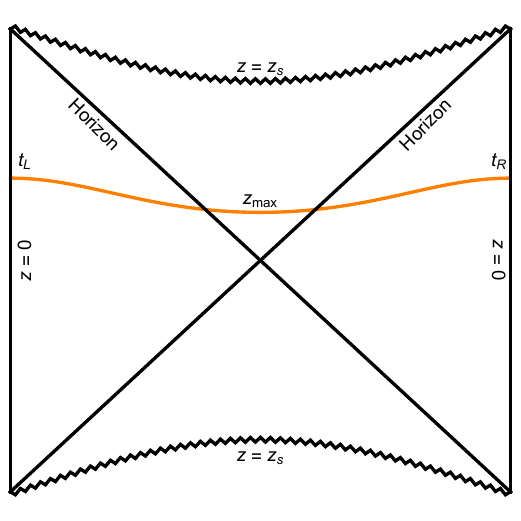}
\caption{Penrose diagram for the extremal surface (orange curve) passing through the horizon. Here $z=z_s$ denotes black hole singularity. We have $z_s=\infty$ for $\lambda_{\text{GB}}\ge 0$ and $ z_s= \left(1-\frac{1}{4\lambda_{\text{GB}}}\right)^{{1}/{d}}z_h$ for $\lambda_{\text{GB}}<0$. 
$z_{\max}$ is the turning point of extremal surface. }
\label{penrose}
\end{figure}

\begin{figure}[t]
\centering
\includegraphics[width=10cm]{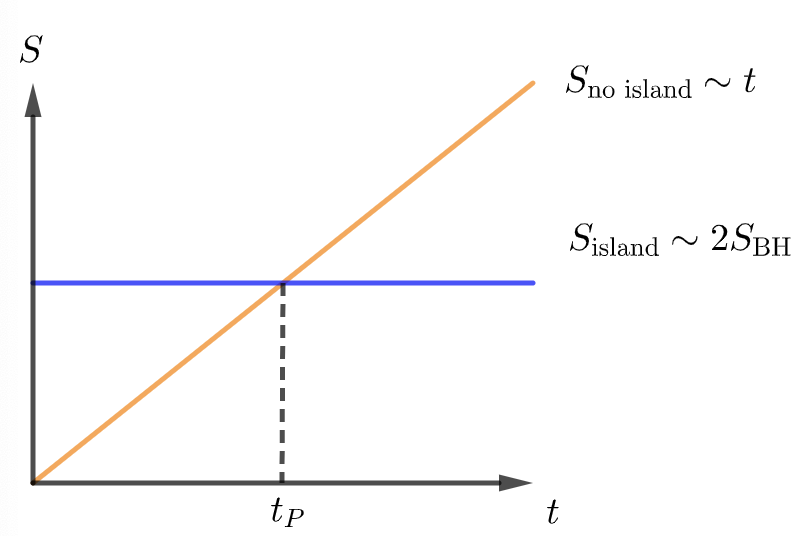}
\caption{Page curve for eternal two-sided black hole is combined by the orange curve when $t< t_P$ and the blue curve when $t\ge t_P$, where $t_P$ is the Page time. Without island, entanglement entropy of Hawking radiation will grow linearly in time after a few thermal times (orange curve). It will exceed twice the black hole entropy and lead to a paradox. The paradox can be resolved by the blue curve (island), which is due to the extremal surface outside the horizon. }
\label{pagecurve}
\end{figure}

Now let us discuss the resolution to the above paradox in the so-called doubly-holographic setup.  See Fig.\ref{doubleholo} for the geometry, where the eternal black hole lives on the end-of-the-world brane $Q$, and the bath lives on the AdS boundary $M$. The entanglement entropy between the black hole and the bath can be calculated 
holographically by holographic entanglement entropy \cite{Ryu:2006bv} in the bulk. 
In general, there are two kinds of extremal surfaces in the bulk. See Fig.\ref{doubleholo} for example. The first kind (orange curve) passes through the horizon and its area increases linearly in time after a few thermal times, due to the stretching of space inside the black hole horizon \cite{Almheiri:2019yqk,Almheiri:2019psy}.  
See also Fig.\ref{penrose} for the  first kind of extremal surface (orange curve) in Penrose diagram. 
While the second kind of extremal surface (blue curve) ending on the brane is independent of time, due to the static nature of spacetime outside the horizon.  The first kind of extremal surface (orange curve) has smaller area at the beginning. Thus it is dominant at early times. As we will show below, the holographic entanglement entropy associated with it increases linearly in time and will exceed the black hole entropy at late times. This is the information paradox for eternal black holes discussed above.  Thanks to the existence of the second kind of extremal surface (blue curve) ending on the brane, it becomes dominant at late times since the entanglement entropy associated with it is a constant.  As a result, the entanglement entropy will firstly increase linearly in time and then saturate a constant value at late enough time.  In summary, due to the island, the Page curve can be recovered so that the information paradox for eternal black holes can be resolved. See Fig.\ref{pagecurve} for the schematic diagram of Page curve for eternal two-sided black hole.

Let us begin to study the information paradox in AdS/BCFT for Gauss-Bonnet gravity.  For the action (\ref{GBactionbulk}), the  black hole solution is given by \cite{Buchel:2009sk}
\begin{eqnarray} \label{GBBH}
ds^2 = \frac{r^2}{L^2} \left(-\frac{f(r)}{f_\infty} dt^2 + dx^2 + \sum_{a=1}^{d-2}\left(dy^a \right)^2 \right)  +\frac{L^2}{r^2} \frac{dr^2}{f(r)},
\end{eqnarray}
with
\begin{eqnarray} \label{island:fr}
&&f(r) = \frac{1}{2\lambda_{\text{GB}}} \left(1-\sqrt{1-4\lambda_\text{GB}\left(1-\frac{r^{d}_h}{r^{d}} \right)} \right),\\
&&f_{\infty}= \lim_{r\rightarrow \infty} f(r) = \frac{1-\sqrt{1-4\lambda_\text{GB}}}{2\lambda_\text{GB}}.  \label{island:finfty}
\end{eqnarray}
The factor $f_\infty$ is chosen so that the speed of light is one on the AdS boundary, i.e., $\lim_{r\rightarrow\infty}g_{tt}/g_{yy}=-1$  \cite{Buchel:2009sk}. Following the previous sections, we set the AdS curvature scale to be one, i.e., $l=L/\sqrt{f_\infty}=1$. The brane $Q$ is located at $x=0$.  Note that (\ref{GBBH}) is a solution to AdS/BCFT if and only if the brane tension vanishes, i.e., $T=\rho=0$.  

By performing the coordinate transformation $z=L/r$, we rewrite the metric (\ref{GBBH}) in Poincar\'e coordinate
\begin{eqnarray} \label{GBBH1}
ds^2 = \frac{1}{ z^2}\left(-\frac{f(z)}{f_\infty}dt^2 +\frac{f_\infty}{f(z)} dz^2 + dx^2+ \sum_{a=1}^{d-2}\left(dy^a \right)^2  \right),
\end{eqnarray}
where
\begin{eqnarray} \label{island:fz}
   f(z)=\frac{1}{2\lambda_\text{GB}} \left(1-\sqrt{1-4\lambda_\text{GB}\left(1-\frac{z^{d}}{z^{d}_h} \right)} \right).
\end{eqnarray}
To study the extremal surface inside the horizon, it is convenient to use the infalling Eddington-Finkelstein coordinates 
\begin{eqnarray} \label{island:coordinatev}
v=t-\int \frac{f_{\infty}}{f(z)} dz. 
\end{eqnarray}
Then the metric (\ref{GBBH1}) becomes
\begin{eqnarray} \label{GBBH2}
ds^2 = \frac{1}{ z^2}\left(-\frac{f(z)}{f_\infty}dv^2 -2dvdz + dx^2+ \sum_{a=1}^{d-2}\left(dy^a \right)^2  \right).
\end{eqnarray}

The holographic entanglement entropy for Gauss-Bonnet gravity is given by \cite{Hung:2011xb},
\begin{eqnarray} \label{island:HEE}
S = \frac{1}{4 G_N} \int_m d^{d-1}x\sqrt{\gamma}\left(1+\frac{2L^2 \lambda_{\text{GB}}}{(d-2)(d-3)} \mathcal{R} \right) +   \frac{1}{ G_N} \int_{\partial m} d^{d-2}x\sqrt{\sigma} \frac{L^2 \lambda_{\text{GB}}}{(d-2)(d-3)}\mathcal{K}  .
\end{eqnarray}
where $m$ denotes the codim-2 extremal surface,  $\mathcal{R}$ is the intrinsic Ricci scalar on $m$, $\partial m$ is the boundary of $m$ and  $\mathcal{K} $ is the extrinsic curvature on $\partial m$. 

\subsection{ No-island phase}

As we have mentioned above, there are two kinds of extremal surfaces in AdS/BCFT, and the holographic entanglement entropy is related to the one with smaller `area'.  Here `area' denotes the area with corrections (\ref{island:HEE})  from Gauss-Bonnet terms.  In this subsection, we focus on the extremal surface passing through the horizon, which is labelled by the orange curve in 
Fig.\ref{doubleholo} and Fig.\ref{penrose}.  We denote the entropy functional of the extremal surface passing through the horizon by ``extremal surface `area' " in this subsection. Whether it corresponds to entanglement entropy depends on if it has smaller `area' among all kinds of extremal surfaces.

The embedding function of the extremal surface passing through the horizon is given by
\begin{eqnarray} \label{island:curveI}
{\text{orange curve}}:\   v=v(z), \ \ x=x_0,
\end{eqnarray}
where $x_0$ is a constant. 
Substitute (\ref{island:curveI}) and (\ref{GBBH2}) into (\ref{island:HEE}), we obtain the entropy functional
\begin{eqnarray} \label{island:HEEtypeI}
S =\int dz \mathcal{L}(z, v', v'')= \frac{V_{d-2}}{4 G_N} \int_0^{z_{\max}} dz \sqrt{\frac{-v'(z) \left(2 f_{\infty }+f(z) v'(z)\right)}{z^{2 (d-1)} f_{\infty }}}\left(1+\frac{2f_{\infty} \lambda_{\text{GB}}}{(d-2)(d-3)} \mathcal{R} \right) ,
\end{eqnarray}
where $S$ is half of the 
 extremal surface `area'
 of the two-side black hole, $z=z_{\max}$ is the turning point (see Fig.\ref{penrose}), $V_{d-2}=\int d^{d-2}y$ is the volume of horizontal space and
\begin{eqnarray} \label{island:RtypeI}
\mathcal{R} =\frac{(d-2) f_{\infty } \left(v'(z) \left(v'(z) \left((d-1) f(z)+z f'(z)\right)+2 (d-1) f_{\infty }\right)+2 z v''(z) \left(f(z) v'(z)+f_{\infty }\right)\right)}{v'(z)^2 \left(f(z) v'(z)+2 f_{\infty }\right){}^2},
\end{eqnarray}
depends on $v''(z)$.  
Note that the entropy functional (\ref{island:HEEtypeI}) 
does not depend on $v(z)$ exactly. 
As a result, the Euler-Lagrange equation derived from (\ref{island:HEEtypeI}) becomes
 \begin{eqnarray}\label{onislandELeq}
\frac{d}{dz} \Big( \frac{d}{dz}\frac{\partial \mathcal{L}}{\partial v''(z)}-\frac{\partial \mathcal{L}}{\partial v'(z)} \Big)=0.
\end{eqnarray}
From (\ref{onislandELeq}), we can define a conserved quantity
\begin{eqnarray} \label{island: EI}
E_\textrm{I}&=&\frac{d}{dz} \frac{\partial \mathcal{L}}{\partial v''(z)}-\frac{\partial \mathcal{L}}{\partial v'(z)}\nonumber\\
&=& -\frac{V_{d-2}}{4 G_N}\frac{\left(f(z) v'(z)+f_{\infty }\right) \left(2 \lambda_{\text{GB}}  f_{\infty }^2+2 f_{\infty } v'(z)+f(z) v'(z)^2\right)}{z^{d-1} f_{\infty }^2 \left(-\frac{ v'(z) \left(f(z) v'(z)+2 f_{\infty }\right)}{f_{\infty }}\right){}^{3/2}}.
\end{eqnarray}
By symmetry, we have $v'(z_{\max})=-\infty$ \cite{Carmi:2017jqz} at the turning point $z=z_{\max}$. Substituting $v'(z_{\max})=-\infty$ into the above equation, we  get
\begin{eqnarray} \label{island: EI1}
E_\textrm{I}&=& \frac{V_{d-2}}{4 G_N}z_{\max }^{1-d} \sqrt{-\frac{f\left(z_{\max }\right)}{f_{\infty }}}\nonumber\\
&=&-\frac{V_{d-2}}{4 G_N}\frac{\left(f(z) v'(z)+f_{\infty }\right) \left(2 \lambda_{\text{GB}}   f_{\infty }^2+2 f_{\infty } v'(z)+f(z) v'(z)^2\right)}{z^{ d-1} f_{\infty }^2 \left(-\frac{ v'(z) \left(f(z) v'(z)+2 f_{\infty }\right)}{f_{\infty }}\right){}^{3/2}}.
\end{eqnarray}
From (\ref{island: EI1}), we can solve $v'(z)$ in terms of $z$ and $z_{\text{max}}$ \footnote{It is easier to solve the combination $t'(z)=v'(z)+\frac{f_{\infty}}{f(z)}$ from (\ref{island: EI1}), where $t'(z)^2$ is the root of a cubic equation.}. Then we can obtain the time on the AdS boundary $M$
\begin{eqnarray} \label{island:timeI}
t=t(0)-t(z_{\text{max}})=-\int_{0}^{z_{\text{max}}} t'(z) dz=-\int_0^{z_{\text{max}}} \Big(v'(z)+\frac{f_{\infty}}{f(z)} \Big) dz,
\end{eqnarray}
where $t=t(0)=t_R=t_L$ is the time on the right (left) AdS boundary $M$, $t(z_{\text{max}})=0$ is the time with respect to the turning point \cite{Carmi:2017jqz}.

We are interested in the evolution of 
extremal surface `area'  (\ref{island:HEEtypeI}) at late times.  In the large-time limit, the conserved quantity (\ref{island: EI1}) approaches to an extremum, i.e., $\lim_{t\to \infty} d E_\textrm{I}/ d z_{\max }=0$ \cite{Carmi:2017jqz}, which yields 
\begin{eqnarray} \label{island:zmaxlargetime}
2 (d-1) f\left(\bar{z}_{\max }\right)-\bar{z}_{\max } f'\left(\bar{z}_{\max }\right)=0,
\end{eqnarray}
where $\bar{z}_{\max }=\lim_{t\to \infty} z_{\max }$. Solving the above equation, we get
\begin{eqnarray} \label{island:largetimezmax}
\bar{z}_{\max }=\left(12d-16 -\frac{d-2}{\lambda_\text{GB}}+\frac{\sqrt{(d-2)^2+4d(3d-4)\lambda_\text{GB}}}{\lambda_\text{GB}} \right)^{1/d} \frac{(d-1)^{1/d}}{(3d-4)^{2/d}}   z_h.
\end{eqnarray}
From (\ref{island: EI1}), we can solve $v'(z)$ in terms of $E_\textrm{I}$. Thus, we have 
\begin{eqnarray} \label{island:prove1}
\lim_{t\to \infty}\frac{\partial v'(z)}{ \partial  z_{\max }}=\lim_{t\to \infty} \frac{\partial  v'(z)}{ \partial  E_\textrm{I}} \frac{ \partial  E_\textrm{I}}{ \partial z_{\max } }=0,
\end{eqnarray}
where we have used $\lim_{t\to \infty} d E_\textrm{I}/ d z_{\max }=0$ \cite{Carmi:2017jqz}.   By using EOM \footnote{ The EOM can be derived by taking derivative of (\ref{island: EI1}) with respect to 
$z$.}, we can express $v''(z)$ in function of $z$ and $v'(z)$ so that $\mathcal{L}(z,v',v'')$ becomes a function of only $z$ and $v'(z)$.  As a result, we have 
\begin{eqnarray} \label{island:prove2}
\lim_{t\to \infty}\frac{\partial  \mathcal{L}(z,v',v'')}{ \partial  z_{\max }}=\lim_{t\to \infty} \frac{\partial  \mathcal{L}(z,v',v''(v'))}{ \partial  v'} \frac{ \partial v'}{ \partial z_{\max } }=0. 
\end{eqnarray}

Now we are ready to study the time evolution of 
extremal surface `area'  (\ref{island:HEEtypeI}) 
at late times.  
By applying (\ref{island:HEEtypeI}, \ref{island:timeI}) together with (\ref{island:prove1}, \ref{island:prove2}), we derive
\begin{eqnarray} \label{island:largetime}
\lim_{t\to \infty}\frac{dS}{dt}&=&\lim_{t\to \infty}\frac{dS/dz_{\text{max}}}{dt/dz_{\text{max}}}=\frac{\mathcal{L}\Big(\bar{z}_{\max },v'(\bar{z}_{\max }),v''(\bar{z}_{\max })\Big)+\int_0^{\bar{z}_{\max }} \frac{\partial \mathcal{L}(z,v',v'')}{ \partial  \bar{z}_{\max }} dz }{-v'(\bar{z}_{\max })-\frac{f_{\infty}}{f(\bar{z}_{\max })} -\int_0^{\bar{z}_{\max }} \frac{\partial v'}{ \partial  \bar{z}_{\max }} dz }
\nonumber\\
&=&\frac{\mathcal{L}\Big(\bar{z}_{\max },v'(\bar{z}_{\max }),v''(\bar{z}_{\max })\Big)}{-v'(\bar{z}_{\max })-\frac{f_{\infty}}{f(\bar{z}_{\max })} }\nonumber\\
&=&\frac{d+2\lambda_{\text{GB}}   \bar{z}_{\max } f'\left(\bar{z}_{\max }\right)-8 \lambda_{\text{GB}}   f\left(\bar{z}_{\max }\right)-3}{(d-3) \left(4 \lambda_{\text{GB}}   f\left(\bar{z}_{\max }\right)+1\right)} 
E_\textrm{I},
\end{eqnarray}
where we have used $\bar{z}_{\max }=\lim_{t\to \infty} z_{\max }$, $v'(\bar{z}_{\max })=-\infty$ and EOM to delete $v''(z)$ in $\mathcal{L}(z,v',v'')$. 
Substituting (\ref{island:zmaxlargetime}) into (\ref{island:largetime}), we finally obtain
\begin{eqnarray} \label{island:largetime1}
\lim_{t\to \infty} \frac{dS}{dt}
=\lim_{t\to \infty}  E_\textrm{I}
= \frac{V_{d-2}}{4 G_N}\bar{z}_{\max }^{1-d} \sqrt{-\frac{f\left(\bar{z}_{\max }\right)}{f_{\infty }}} ,
\end{eqnarray}
which is a constant. As expected, the 
extremal surface `area' 
increases linearly with time in late times.  

\begin{figure}[t]
\centering
\includegraphics[width=7.6cm]{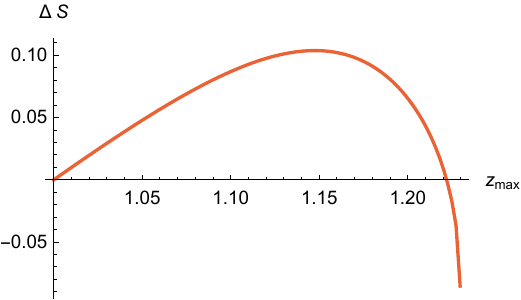}
\includegraphics[width=7.6cm]{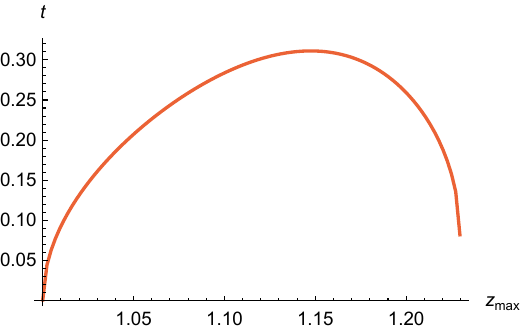}
\caption{ For $d=4$ and $\lambda_{\text{GB}}= -7/36$ which violates the bound (\ref{island:conditionlambda}), the 
extremal surface `area'  $\Delta S$
and the time $t$ are no longer monotonic functions of $z_{\max}$. Instead, they become maximum at $z_{\max}\approx 1.148$. As a result, the time evolution of 
extremal surface `area'  
stops at the finite time $t(1.148)\approx 0.311$. See Fig.\ref{entropytimestop} below.
For simplicity, we set ${V_{d-2}}/({4 G_N}) = 1 $ in this and the following figures.
}
\label{maxentropytime}
\end{figure}

\begin{figure}[t]
\centering
\includegraphics[width=10cm]{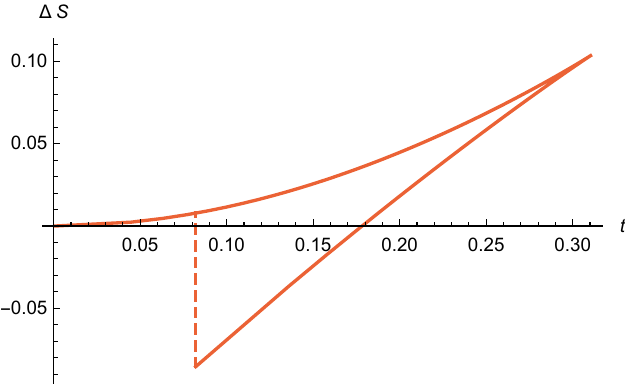}
\caption{ For $d=4$ and $\lambda_{\text{GB}}= -7/36$ which violates the bound (\ref{island:conditionlambda}), the time evolution of 
extremal surface `area'  
stops at a finite time $t\approx 0.311$. Note that we always select the smaller `area'  for entanglement entropy, when one time corresponds to two extremal `areas' in the red polyline. As a result, there is a 
zeroth-order phase transition of entanglement entropy at  $t\approx 0.082$.}
\label{entropytimestop}
\end{figure}

Note that (\ref{island:largetimezmax}) and thus (\ref{island:largetime1}) is well-defined if and only if 
\begin{eqnarray} \label{island:conditionlambda}
 \lambda_{\text{GB}} \ge \lambda_c=-\frac{(d-2)^2}{4d(3d-4)},
\end{eqnarray}
which is within the causality constraint (\ref{GBconstraint}).
If the bound (\ref{island:conditionlambda}) is violated, there is no extremum for the conserved quantity (\ref{island: EI1}) 
\begin{eqnarray} \label{noisland: EInoextremum}
d E_\textrm{I}/ d z_{\max }\ne 0
\end{eqnarray}
and the linear growth rate (\ref{island:largetime1}) in the late time limit breaks down
\begin{eqnarray} \label{noisland: growthbreaksdown}
\lim_{t\to \infty} \frac{dS}{dt}=\lim_{t\to \infty}  E_\textrm{I} \ \text{ becomes complex}.
\end{eqnarray}
This implies that the extremal surface passing through the horizon is not well-defined in the late time limit $t\to \infty$ for $ \lambda_{\text{GB}} < \lambda_c$. In fact, this unusual case also appears in the study of complexity=volume conjecture of holographic complexity \cite{An:2018dbz} \footnote{Eq.(3.17) of \cite{An:2018dbz} yields a similar but different bound $\lambda_{\text{GB}} \ge -1/12$. The different bound is due to the fact that holographic complexity is related to the codimension-one extremal surface, while the holographic entanglement entropy is relevant to the codimension-two extremal surface.}. For our case, numeral calculations indeed confirm that the extremal surface passing through the horizon can not be defined after some finite time. The reasons are as follows. As shown in Fig.\ref{maxentropytime}, when the bound (\ref{island:conditionlambda}) is violated, 
the extremal surface `area'  
and the time associated to the extremal surfaces passing through the horizon are no longer monotonic functions of $z_{\max}$ and there are maximum values for
 them. As a result, the time evolution of 
the extremal surface `area' stops at finite times.  See Fig.(\ref{entropytimestop}).  
 See also appendix B for an analytic understanding of this finite-time problem in a toy model.

This is problematic in AdS/CFT, since this means that the entanglement entropy is not well-defined after some finite time. To avoid this situation, one has to impose the lower bound (\ref{island:conditionlambda}) for the couplings in AdS/CFT.  Interestingly, this lower bound is stronger than the causality constraint (\ref{GBconstraint}). 

However, this is not a problem in AdS/BCFT. Recall that we have two kinds of extremal surfaces in AdS/BCFT: the first kind passes through the horizon and evolves with time, while the second kind ends on the brane and remains invariant. For sufficiently negative $\lambda_{GB}$, after some finite time, the first kind of extremal surface cannot be defined, but the second kind of extremal surface is always well-defined. Thanks to the second kind of extremal surface (island), the entanglement entropy can be defined at any time in AdS/BCFT. 

Note that, for negative $\lambda_{\text{GB}}$, we have
\begin{eqnarray} \label{islandzs}
0 \le z \le z_s= \left(1-\frac{1}{4\lambda_{\text{GB}}}\right)^{{1}/{d}}z_h,
\end{eqnarray}
where $z_s$ denotes the black hole singularity. Note also that, when the bound (\ref{island:conditionlambda}) is violated,  $z_{\max}$ can approach to the singularity $z_s$. Remarkably, the 
extremal surface `area'  
(\ref{island:HEEtypeI}) and the time (\ref{island:timeI}) are both well-defined even for $z_{\max}=z_s$.  Thus we safely have $z_h\le z_{\max}\le z_s$ in Fig. \ref{maxentropytime}. For simplicity, we set $z_h=1$ in all of the figures of this paper.

For sufficiently negative $\lambda_{\text{GB}}$, one time corresponds to two 
extremal surface `area' .
See the red polyline of Fig.(\ref{entropytimestop}). For this case, we always select the smaller
`area' for the entropy.
 As a result, there is a 
zeroth-order phase transition of entanglement entropy. In other words, the time evolution of entanglement entropy is discontinuous.
Take $d=4$ and $\lambda_{\text{GB}}=-7/36$ as an example. During $0< t <  0.082$, $\Delta S=S(t)-S(0)$ is positive and increases monotonically. At time $t\approx 0.082$, $\Delta S$  becomes negative and there is a 
zeroth-order phase transition of entanglement entropy. Then $\Delta S$ continues to increase monotonically until it stops at the maximum time $t_{\text{max}}\approx 0.311$. 
After the maximum time $t>t_{\text{max}}$, there is no well-defined extremal surface inside the horizon.  As we will show in the next subsection, there is a well-defined extremal surface outside the horizon, which is related to the so-called island. 
 We leave a careful study of this 
zeroth-order phase transition of entanglement entropy to future works.  See also \cite{Liu:2012eea} for other kinds of 
zeroth-order phase transitions of entanglement entropy.

\begin{figure}[t]
\centering
\includegraphics[width=7.6cm]{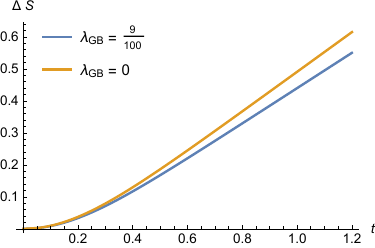}
\includegraphics[width=7.6cm]{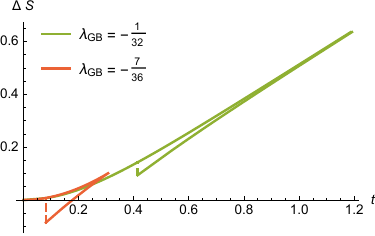}
\caption{Time evolution of 
extremal surface `area'  
associated with the extremal surface passing through the horizon for $d=4$. The 
 extremal surface `area'  
(orange, blue and green lines) increases linearly with time at late times within the bound (\ref{island:conditionlambda}), i.e., $\lambda_{\text{GB}}\geq \lambda_c=  -1/32$. On the other hand, the 
the extremal surface `area'  
(red line) terminates growth and the extremal surface inside the horizon disappears after a finite time if the bound (\ref{island:conditionlambda}) is violated, i.e., $\lambda_{\text{GB}}= -7/36 <  \lambda_c $. Note that there is a 
zeroth-order phase transition of entanglement entropy at early times for the green line and red line. }
\label{entropytypeI}
\end{figure}

To end this section, let us discuss the time evolution of 
the extremal surface `area'  
associated to the extremal surfaces passing through the horizon for various couplings $\lambda_{\text{GB}} $. See Fig.\ref{entropytypeI}, where $\Delta S=S(t)-S(0)$, $S$ denotes half of the total 
extremal surface `area' of two-side black hole and $t=t_R=t_L$ is half of the total time $t_R+t_L$. Here we only summarize the results and leave the derivations to Appendix B.  For simplicity, we focus on $d=4$. Note that the properties of the conserved quantity $E_\textrm{I}$ (\ref{island: EI1}) are quite different in different ranges of couplings $\lambda_{\text{GB}} $. And this leads to various novel phase transitions. There are three kinds of conserved quantity $E_\textrm{I}$. In the normal case including Einstein gravity
\begin{eqnarray} \label{EIcase1}
\text{Case I}:\ - \frac{18\sqrt{3}-31}{16}\le \lambda_{\text{GB}}, 
\end{eqnarray}
the local extremal point $\bar{z}_{\text{max}}$ is also the global maximum point for the conserved quantity $E_\textrm{I}$ (\ref{island: EI1}). In this case,  
the extremal surface `area'  
is a monotonic single-valued function of the time.  See the orange and blue curves of Fig.\ref{entropytypeI}. In the second case 
\begin{eqnarray} \label{EIcase2}
\text{Case II}:\ -1/32=\lambda_c\le \lambda_{\text{GB}}<- \frac{18\sqrt{3}-31}{16}, 
\end{eqnarray}
 the local extremal point $\bar{z}_{\text{max}}$ (\ref{island:largetimezmax}) is no longer the global maximum point for the conserved quantity (\ref{island: EI1}). As a result, 
the extremal surface `area'  
 is no longer a single-valued function of the time.  See the green curve of Fig.\ref{entropytypeI}.  In the third case 
 \begin{eqnarray} \label{EIcase3}
\text{Case III}:\ \lambda_{\text{GB}}<\lambda_c=-1/32, 
\end{eqnarray}
there is no local extremal point $\bar{z}_{\text{max}}$ for the conserved quantity (\ref{island: EI1}).  As we have discussed above, the first kind of extremal surfaces can not be defined after some finite time for this case.  At early times, similar to the second case, 
the extremal surface `area'  
 is not a single-valued function of the time either.  See the red curve of Fig.\ref{entropytypeI}.  Note that one time corresponds to 
two extremal surface `area'  
 in case II and case III. For these cases, we always select the smaller one to be the entanglement entropy. As a result, 
there are zeroth-order phase transitions of entanglement entropy in these two cases. 

\subsection{ Island phase}

Let us go on to discuss the extremal surface ending on the brane, which is related to the so-called island.  See the blue curve of Fig.\ref{doubleholo}.  The embedding function of the extremal surface is given by
\begin{eqnarray} \label{island:curveII}
{\text{blue curve}}:  x=x(z), \ t=\text{constant}.
\end{eqnarray}
Substitute (\ref{island:curveII}) and (\ref{GBBH1}) into (\ref{island:HEE}), we obtain 
\begin{eqnarray} \label{island:HEEtypeII}
S =\int dz \mathcal{L}(z, x', x'')= \frac{V_{d-2}}{4 G_N} \int_0^{z_0} dz \sqrt{\frac{f(z) x'(z)^2+f_{\infty }}{z^{2 (d-1)} f(z)}}\left(1+\frac{2f_{\infty} \lambda_{\text{GB}}}{(d-2)(d-3)} \mathcal{R} \right) ,
\end{eqnarray}
where $z_0$ is location of the island on the brane (blue point of Fig.\ref{doubleholo} ) and
\begin{eqnarray} \label{island:RtypeII}
\mathcal{R} =\frac{(d-2) \left(-f(z)^2 x'(z) \left((d-1) x'(z)+2 z x''(z)\right)-(d-1) f(z) f_{\infty }+z f_{\infty } f'(z)\right)}{\left(f(z) x'(z)^2+f_{\infty }\right){}^2}.
\end{eqnarray}
Note that the entropy functional (\ref{island:HEEtypeII})
does not depend on $x(z)$ exactly. As a result, we can define a conserved quantity
\begin{eqnarray} \label{island: EII}
E_\textrm{II}&=&\frac{d}{dz} \frac{\partial \mathcal{L}}{\partial x''(z)}-\frac{\partial \mathcal{L}}{\partial x'(z)}\nonumber\\
&=&- \frac{V_{d-2}}{4 G_N}\frac{z^{1- d} x'(z) \left(f(z) x'(z)^2+f_{\infty } (1-2 \lambda_{\text{GB}}  f(z))\right)}{f(z) \left(\frac{ f(z) x'(z)^2+f_{\infty }}{f(z)}\right){}^{3/2}}.
\end{eqnarray}
From the blue curve of Fig.\ref{doubleholo}, we observe that $x'(z_0)=-\infty$. Substituting $x'(z_0)=-\infty$ into the above equation, we  get
\begin{eqnarray} \label{island: EII1}
E_\textrm{II}&=& \frac{V_{d-2}}{4 G_N}z_{0}^{1-d}\nonumber\\
&=& - \frac{V_{d-2}}{4 G_N}\frac{z^{1- d} x'(z) \left(f(z) x'(z)^2+f_{\infty } (1-2 \lambda_{\text{GB}}  f(z))\right)}{f(z) \left(\frac{f(z) x'(z)^2+f_{\infty }}{f(z)}\right){}^{3/2}}.
\end{eqnarray}
From (\ref{island: EII1}), we can solve $x'(z)$ in terms of $z$ and $z_0$. We choose the negative root 
\begin{eqnarray} \label{island: xsolution}
x'(z)=-\sqrt{X},
\end{eqnarray}
where
\begin{eqnarray} \label{island: xsolution1}
&&X=-\frac{c_1}{3}+\frac{\sqrt[3]{-2 c_1^3+9 c_2 c_1+\sqrt{\left(2 c_1^3-9 c_2 c_1+27 c_3\right){}^2-4 \left(c_1^2-3 c_2\right){}^3}-27 c_3}}{3 \sqrt[3]{2}}\nonumber\\
&&\ \ \ \ \ \ +\frac{\sqrt[3]{2} \left(c_1^2-3 c_2\right)}{3 \sqrt[3]{-2 c_1^3+9 c_2 c_1+\sqrt{\left(2 c_1^3-9 c_2 c_1+27 c_3\right){}^2-4 \left(c_1^2-3 c_2\right){}^3}-27 c_3}},
\end{eqnarray} 
is a root of a cubic equation with the parameters
\begin{eqnarray} \label{island: xsolution2}
&&c_1=-\frac{f_{\infty } \left(2 z^2 z_0^{2 d} (2 \lambda_{\text{GB}}  f(z)-1)+3 z_0^2 z^{2 d}\right)}{\left(z^2 z_0^{2 d}-z_0^2 z^{2 d}\right) f(z)},\nonumber\\
&&c_2=\frac{f_{\infty }^2 \left(z^2 (1-2 \lambda_{\text{GB}}  f(z))^2-3 z^{2 d} z_0^{2-2 d}\right)}{\left(z^2-z^{2 d} z_0^{2-2 d}\right) f(z)^2},\nonumber\\
&&c_3=\frac{z_0^2 z^{2 d} f_{\infty }^3}{\left(z_0^2 z^{2 d}-z^2 z_0^{2 d}\right) f(z)^3}.
\end{eqnarray} 
By using (\ref{island: xsolution}), we can derive the entropy (\ref{island:HEEtypeII}) and the location of the endpoint of the extremal surface on the AdS boundary $M$
\begin{eqnarray} \label{island:xII}
x_0=x(0)=-\int_0^{z_{0}} x'(z) dz,
\end{eqnarray}
where we have used $x(z_0)=0$.  Note that the entanglement entropy  (\ref{island:HEEtypeII}) is divergent due to the integral near the AdS boundary $z\sim 0$. We are interested in the entropy difference between the two kinds of extremal surfaces (blue curve and orange curve of Fig.\ref{doubleholo} ). It turns out that the entropy difference $\Delta S=S-S_I(t=0)$ is finite \footnote{That is because the divergent terms of entanglement entropy are independent of the states of CFTs, thus they can be cancelled by the entropy difference.}, 
where $S_I(t=0)$ denotes the entropy at the initial moment for the extremal surface (orange curve of Fig.\ref{doubleholo}) passing through the horizon.

\begin{figure}[t]
\centering
\includegraphics[width=7.6cm]{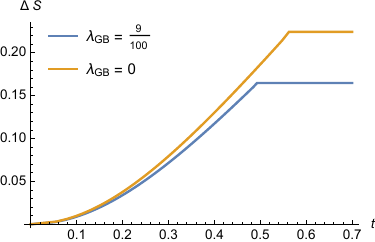}
\includegraphics[width=7.6cm]{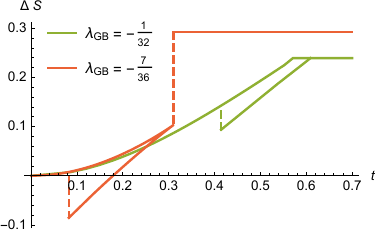}
\caption{Three kinds of Page curve for eternal Gauss-Bonnet black holes with $d=4$. Case I (orange and blue lines obeying bound (\ref{island:conditionlambda})): the entanglement entropy keeps increasing until it saturates a constant value. Case II (green line obeying bound (\ref{island:conditionlambda})): the entanglement entropy firstly increases, then drops suddenly and continues to increase until it saturates a constant value.  Case III (red line violating bound (\ref{island:conditionlambda})): the time evolution of entanglement entropy is discontinuous and there are two 
zeroth-order phase transitions. Note that we always select the smaller entropy, when one time corresponds to two 
extremal surface `area'  
in the above figure. }
\label{entropytypeII}
\end{figure}

As we have discussed in the above and this subsections, there are two choices for the holographic entanglement entropy. The first one (orange curve of Fig.\ref{doubleholo}) depends on time and is smaller at early times, while the second one (blue curve of Fig.\ref{doubleholo}) is a constant and is smaller at late times. 
By definitions, the holographic entanglement entropy is given by the minimal one. In this way, we recover the Page curve for the eternal  black hole. As shown in Fig.\ref{entropytypeII}, there are three kinds of Page curve for Gauss-Bonnet black holes with $d=4$. In the normal case (orange and blue lines of Fig.\ref{entropytypeII}), 
\begin{eqnarray} \label{island:case1}
- \frac{18\sqrt{3}-31}{16}\le \lambda_{\text{GB}}\le \frac{9}{100},
\end{eqnarray}
the entanglement entropy keeps increasing until it saturates a constant value. There is a 
first-order phase transition at the Page time. In the second case  (green line of Fig.\ref{entropytypeII}), 
\begin{eqnarray} \label{island:case2}
-\frac{1}{32}\le \lambda_{\text{GB}}< - \frac{18\sqrt{3}-31}{16},
\end{eqnarray}
the entanglement entropy firstly increases, then drops suddenly and continues to increase until it saturates a constant value. There is 
a zeroth-order phase transition at the early time and a first-order phase transition at the late time for the second case. In the third case (red line of Fig.\ref{entropytypeII}), 
\begin{eqnarray} \label{island:case3}
-\frac{7}{36}\le \lambda_{\text{GB}}< -\frac{1}{32},
\end{eqnarray}
there are two zeroth-order phase transitions of entanglement entropy at $t\approx 0.082$ and $t\approx 0.311$, respectively.  At late times, the entanglement entropy also becomes a constant.

 To end this section, let us make some comments.  {\bf 1}. As shown in Fig.\ref{entropytypeII} and appendix B, there are two kinds of phase-transition points for the Gauss-Bonnet couplings, which are given by $\lambda_c$ (\ref{island:conditionlambda}) and $\lambda_f$ (\ref{appB:lambdaGB1}), respectively.   For $ \lambda_c\le \lambda_{\text{GB}}<\lambda_f$, the local extremal point $\bar{z}_{\text{max}}$ (\ref{island:largetimezmax})  is not the global maximum point for the conserved quantity (\ref{island: EI1}).  And there is a zeroth-order phase transition of entanglement entropy. While for $ \lambda_{\text{GB}}<\lambda_c$, there is no local extremal point for the conserved quantity (\ref{island: EI1}).  As a result, there are two zeroth-order phase transitions of Page curve.  {\bf 2}. So far we keep the location of the island $z_0$ (endpoint of the blue curve on the brane $Q$ in Fig.\ref{doubleholo}) as a free parameter. $z_0$ can be fixed by adding a DGP term $R_Q$ or CFTs on the brane $Q$ and then minimizing the holographic entanglement entropy outside the horizon.  See \cite{Ling:2020laa} for an example. {\bf 3}. The black hole solutions to Einstein gravity are also solutions to the curvature-squared gravity (\ref{Ibulkcurvature}) with $\alpha=0$.  Furthermore, the RT surface of Einstein gravity also extremalizes the holographic entanglement entropy for this kind of curvature-squared gravity. As a result, all of the discussions of the island for Einstein gravity in the literature can be generalized to this kind of curvature-squared gravity. {\bf 4}. For general curvature-squared gravity, the exact black hole solutions are non-known. One may consider a locally AdS space with a hyperbolic black hole on the brane \cite{Chen:2020hmv}. We leave the study of this case to future works.

\section{Conclusions and Discussions}

In this paper, we have investigated the AdS/BCFT for curvature-squared gravity. We find that, in general, the shape of the brane is overly constrained by NBC, i.e., $K_{\mu\nu}=0$. As a result, the dual BCFTs have trivial boundary entropy and A-type boundary central charges. To get a non-trivial  AdS/BCFT for curvature-squared gravity, we propose to impose Dirichlet boundary condition for the massive graviton, while imposing Neumann boundary condition for the massless graviton. In this way, we resolve the overly restrictive problem and obtain non-trivial shape dependence of the one point function of stress tensor. In particular, we find that the B-type boundary central charge, which is related to the norm of displacement operator, is positive. Furthermore, we prove that the holographic g-theorem is obeyed by general curvature-squared gravity. This is a strong support for our results. As a by-product, we derive the formula 
of boundary entropy 
and A-type boundary central charges in general dimensions.  We also study the island for AdS/BCFT of Gauss-Bonnet gravity and verify that the Page curve can be recovered for eternal  black holes. Remarkably, there are 
zeroth-order phase transitions for the entanglement entropy within one range of couplings obeying the causality constraint, which is quite different from the case of Einstein gravity.  
We leave the further study of the novel zeroth-order phase transition of the Page curve to future works.

In this paper, we mainly focus on AdS/BCFT. Now let us make some comments on AdS/CFT.  As we have discussed  in sect.4.1, the extremal surfaces passing through the horizon cannot be defined after some finite time for sufficiently negative couplings $\lambda_{\text{GB}}$. Unlike AdS/BCFT, there is only one kind of extremal surfaces in AdS/CFT. This means that the entanglement entropy is not well-defined after some finite time for sufficiently negative couplings $\lambda_{\text{GB}}$. To rule out this unphysical case, one has to impose a lower bound on the couplings
\begin{eqnarray} \label{AdSCFTbound}
-\frac{(d-2)^2}{4d(3d-4)}\le  \lambda_{\text{GB}}.
\end{eqnarray}
Interestingly, the constraint (\ref{AdSCFTbound}) is stronger than the causality constraint (\ref{GBconstraint}).

Usually, things inconsistent in AdS/CFT continue to be inconsistent in AdS/BCFT. One typical example is the causal constraint of 
$\lambda_{\text{GB}}$
(\ref{GBconstraint}) derived from three point functions of AdS/CFT \cite{Buchel:2009sk}. Consider the regions far away from the boundary, the three point functions of AdS/BCFT is well approximated by those of AdS/CFT. Thus the causal constraint (\ref{GBconstraint}) derived by AdS/CFT should also be obeyed by AdS/BCFT. Otherwise, negative energy fluxes appear.  

On the other hand, there are things which are inconsistent in AdS/CFT becoming consistent in AdS/BCFT, due to the novel effects of the boundary. An interesting example is the bulk solution for gravity coupled with a scalar field, which is singular in AdS/CFT but is well-defined in AdS/BCFT. Please see section 6.2 of \cite{Izumi:2022opi} for example. For the convenience of readers, we show the key points below. Consider the Janus type ansatz for $d>2$
\begin{eqnarray}\label{bulkmetric}
&&ds^2=dr^2+ f(r) ds^2_{AdS_d}, \\
&& \phi=\phi(r). \label{bulkscalar}
\end{eqnarray}
The solution is singular in the sense that there is a naked singularity at $r=r_s$ where $ f(r_s)=0$ and $\phi(r_s)$ diverges. Such a naked singularity must be prohibited for the usual AdS/CFT without an end-of-the-world (EOW) brane. However, in the context of the AdS/BCFT setup, this is not a problem but just means that one has to place the EOW brane before this singularity. As a result, there is no singularity in the bulk of AdS/BCFT. 

Our case is another example that the thing inconsistent in AdS/CFT becomes consistent in AdS/BCFT.   Although the extremal surface passing through the horizon cannot be defined after some finite time when the bound (\ref{AdSCFTbound}) is violated, the entanglement entropy is always well-defined due to the new extremal surface (island) ending on the brane in AdS/BCFT.

 The discussions of sect.4.1 can be naturally generalized to the holographic complexity \cite{Susskind:2014rva,Stanford:2014jda,Brown:2015bva}.  The only difference is that the holographic entanglement entropy  is related to the minimal codimension-two extremal surface, while the holographic complexity \cite{Susskind:2014rva,Stanford:2014jda} is relevant to the maximal codimension-one extremal surface.
By using the complexity=volume conjecture \cite{Susskind:2014rva,Stanford:2014jda}, we derive a new bound for the Gauss-Bonnet couplings 
\begin{eqnarray} \label{complexitybound}
 -\frac{1}{12} \le \lambda_{\text{GB}},
\end{eqnarray}
in order that the time evolution of complexity is well-behaved at late times. 
Note that (\ref{complexitybound}) is weaker than the constraint (\ref{AdSCFTbound}) but stronger than the causality constraint (\ref{GBconstraint}).
Note also that there is no island-like extremal volume outside the horizon in AdS/CFT. As a result, the holographic complexity is not well-defined after a finite time when the bound (\ref{complexitybound}) is violated. To rule out this case, we impose the bound (\ref{complexitybound}).  
It should be stressed that, unlike the entanglement entropy, there is no zeroth-order phase transition for the holographic complexity. That is because we choose the maximal extremal volume rather than the minimal one for the holographic complexity. As a result, although we also have the polyline similar to the green line of  Fig.\ref{entropytypeI} in the time evolution of holographic complexity for one range of couplings, there is no zeroth-order phase transition. 
 It should be mentioned that the holographic complexity for Gauss-Bonnet has been studied in \cite{Cai:2016xho,An:2018dbz}. For simplicity,  they did not discuss the case which violates the bound (\ref{complexitybound}). It is interesting to generalize the discussions of this paper to general higher derivative gravity. It is also interesting to study wedge/cone holography for higher derivative gravity.  We hope these problems could be addressed in the future.

\section*{Acknowledgements}
We thank Yan Liu, Jie Ren, Run-Qiu Yang and Peng-Ju Hu for valuable discussions and comments. 
This work is supported by the National Natural Science Foundation of China (Grant No.11905297) and Guangdong Basic and Applied Basic Research Foundation (No.2020A1515010900).

\appendix

\section{A-type boundary central charge}

In this appendix, we derive the A-type boundary central charge from the holographic Weyl anomaly for general curvature-squared gravity. For simplicity, we mainly focus on the model I of sect.3.2, which agrees with the standard theory of Gauss-Bonnet gravity. We briefly comment  on the case of model II of sect.3.2 at the end of this appendix.

For BCFTs in odd dimensions ($d$ odd), there are non-trivial boundary contributions to Weyl anomaly.  In a $d$-dimensional ball, the boundary Weyl anomaly takes the form 
\begin{eqnarray}\label{Weylanomaly}
\mathcal{A}=-2(-1)^{\frac{d-1}{2}} c_{\text{bdy}} \int_P d^{d-1} y\sqrt{|\sigma|} E_{d-1},
\end{eqnarray}
where $c_{\text{bdy}}$ is the A-type boundary central charge which obeys g-theorem, $\sigma_{ab}$ is the induced metric on the boundary $P$, and $E_{d-1}$ is the Euler density normalized by
\begin{eqnarray}\label{Weylanomaly1}
\int_{S_{d-1}} d^{d-1} y\sqrt{|\sigma|} E_{d-1}=2,
\end{eqnarray}
with $S_{d-1}$ a $(d-1)$-dimensional unit sphere. 

The holographic dual of a ball is given by \cite{Takayanagi:2011zk}
\begin{eqnarray}\label{app:ball}
ds^2=\frac{dz^2+dr^2+r^2 d^2{\Omega_{d-1}}}{z^2},
\end{eqnarray}
where the end-of-the-world brane is located at
\begin{eqnarray}\label{app:Q}
r(z)=\sqrt{r_b^2 \cosh ^2(\rho )-\left(z-r_b \sinh (\rho )\right){}^2},
\end{eqnarray}
with $r_b$ the radius of the ball. Substituting (\ref{app:ball},\ref{app:Q}) into the action of higher curvature gravity (\ref{Ibulkcurvature},\ref{Ibdycurvature}) and integrating along $r$, we get
\begin{eqnarray}\label{app:action}
&&I=-I_E=S(d-1)\int_{\e} dz \Big( -\frac{2 r(z)^d}{z^{d+1}}\nonumber\\
&&\ \ \ \ \ \ \ \ \ \  \ \ +\frac{\tanh (\rho ) \text{sech}(\rho ) \left(-8 \alpha  \left(d^2-3 d+2\right)+\cosh (2 \rho )+1\right)r_b r(z)^{d-2}}{z^d}\Big),
\end{eqnarray}
where $I_E$ is the Euclidean action, $S(d-1)=2 \pi ^{d/2}/\Gamma \left(d/2\right)$ is the volume of $(d-1)$-dimensional unit sphere. 
In the above calculations, we have use $T$ (\ref{GBtension}), $\Psi_{ij}=\bar{R}_{\mu\nu\rho\sigma}=0$ and the following formulas
 \begin{eqnarray}\label{app:formula1}
&&R+d(d-1)=-2d,\  \mathcal{L}_{\text{GB}}(\bar{R})= \bar{R}^2= \bar{R}_{\mu\nu}\bar{R}^{\mu\nu}=0,\\ \label{app:formula2}
&&K_{ij}=\tanh(\rho) h_{ij},\ K=d \tanh(\rho), \ G_Q^{ij}=\frac{1}{2} \left(d^2-3 d+2\right)\text{sech}^2(\rho ) h^{ij},\\ \label{app:formula3}
&& J=-\frac{1}{3} d \left(d^2-3 d+2\right) \tanh ^3(\rho ),\ G^{ij}_{Q} K_{ij}=\frac{1}{2} d \left(d^2-3 d+2\right) \tanh (\rho ) \text{sech}^2(\rho ),
\end{eqnarray}
and 
 \begin{eqnarray}\label{app:formula4}
&& \int_N d^{d+1}x\sqrt{|g|}= S(d-1) \int dz \int_{0}^{r(z)} dr\frac{r^{d-1}}{z^{d+1}}=S(d-1) \int dz \frac{r(z)^{d}}{d\ z^{d+1}},\\ \label{app:formula5}
&&\int_Q d^{d}y\sqrt{|h|} =S(d-1) \int dz \frac{r(z)^{d-1} \sqrt{1+r'(z)^2}}{z^d}=S(d-1) \int dz \frac{r_b \cosh(\rho) r(z)^{d-2} }{z^d}.\nonumber\\
\end{eqnarray}
We verify that (\ref{app:action}) agrees with the results of \cite{Fujita:2011fp} for Einstein gravity \footnote{Note that  \cite{Fujita:2011fp} focus on the Euclidean action $I_E$. Thus $I_E$ of \cite{Fujita:2011fp}  is equal to $-I$ of this paper.}.

According to \cite{Henningson:1998gx}, holographic Weyl anomaly can be obtained from the UV logarithmic divergent term of the gravitational action
\begin{eqnarray}\label{app:Weylanomalytoaction}
\mathcal{A}=I |_{\log \frac{1}{\epsilon} }.
\end{eqnarray}
From (\ref{Weylanomaly}, \ref{Weylanomaly1},\ref{app:Weylanomalytoaction}), we get
\begin{eqnarray}\label{app:cbdyformula}
c_{\text{bdy}}=\frac{1}{4}(-1)^{\frac{d+1}{2}}I |_{\log \frac{1}{\epsilon} }.
\end{eqnarray}
Thus, we only need to consider the coefficient of $1/z$ in the integration (\ref{app:action}). However, this is not an easy task for general odd $d$. Let us first study the case of small $\rho$. We have
\begin{eqnarray}\label{app:actionsmallrho}
&&I=S(d-1)\int_{\e} dz \Big[ -2 z^{-d-1} \left(r_b^2-z^2\right){}^{d/2}\nonumber\\
&&\ \ \ \ \ \ \ \ \ \  -2 (d-1) r_b (4 \alpha  (d-2)+1) z^{-d} \left(r_b^2-z^2\right){}^{\frac{d-2}{2}}\ \rho  \nonumber\\
&&\ \ \ \ \ \ \ \ \ \ -(d-2)^2 r_b^2 (8 \alpha  (d-1)+1) z^{1-d} \left(r_b^2-z^2\right){}^{\frac{d-4}{2}}\  \rho ^2+O\left(\rho ^3\right)
\Big].
\end{eqnarray}
Note that $z^{-d-1} \left(r_b^2-z^2\right){}^{d/2}$ and $z^{1-d} \left(r_b^2-z^2\right){}^{\frac{d-4}{2}}$ contain no $1/z$ terms for odd $d$, and 
\begin{eqnarray}\label{app:formula6log}
z^{-d} \left(r_b^2-z^2\right){}^{\frac{d-2}{2}}|_{1/z}= \frac{(-1)^{\frac{d-1}{2}}\Gamma \left(\frac{d}{2}\right)}{\sqrt{\pi } \Gamma \left(\frac{d+1}{2}\right)} \frac{1}{r_b},
\end{eqnarray}
where ${ }|_{1/z}$ denotes the coefficient of $1/z$ term.
From (\ref{app:cbdyformula}, \ref{app:actionsmallrho},\ref{app:formula6log}), we obtain the A-type boundary central charge
\begin{eqnarray}\label{app:cbdysmallrho}
c_{\text{bdy}}=(1+4 \alpha  (d-2)) S(d-2)\rho+O\left(\rho ^3\right),
\end{eqnarray}
which agrees with the g-function (\ref{app:cbdy123}). 

Let us go on to discuss the case of finite $\rho$. For any given odd $d$, we can derive the holographic Weyl anomaly straightforwardly. We have done calculations up to $d=19$ and find that the A-type boundary central charge takes 
the following expression
\begin{eqnarray}\label{app:cbdy}
c_{\text{bdy}}&=&\frac{S(d-2)\sinh (\rho )}{d-2} \Big[\left(1+4 \alpha  (d-2)^2\right) \cosh ^{d-3}(\rho )\nonumber\\
&&\ \ \ \ \ \ \ \ \ \ \ \ \ \ \ \ \ \ \ \ \ \ \ \ +(d-3) \, _2F_1\left(\frac{1}{2},\frac{5-d}{2};\frac{3}{2};-\sinh ^2(\rho )\right)\Big],
\end{eqnarray}
which is exactly the same as the g-function (\ref{app:cbdy123})
and the universal term of boundary entropy (\ref{sect2.2:bdyentropy3}). This is a strong support for the holographic g-theorem of higher curvature gravity.  To the best of our knowledge, it is the first time that the A-type boundary central charge (\ref{app:cbdy}) is obtained in general dimensions. 

In the above discussions, we focus on model I of sect.3.2. Recall that, on an AdS background, the action of model II of sect.3.3 is exactly the same as the action of Einstein gravity. As a result, the A-type boundary central charge of model II is the same as that of Einstein gravity, which can be obtained from (\ref{app:cbdy}) by taking the limit $\a \to 0$.  Now we finish the derivation of A-type boundary central charges for general curvature-squared gravity.

\section{Time evolution of entanglement entropy for various couplings}


\begin{figure}[!ht]
  \centering
  \includegraphics[width=12cm]{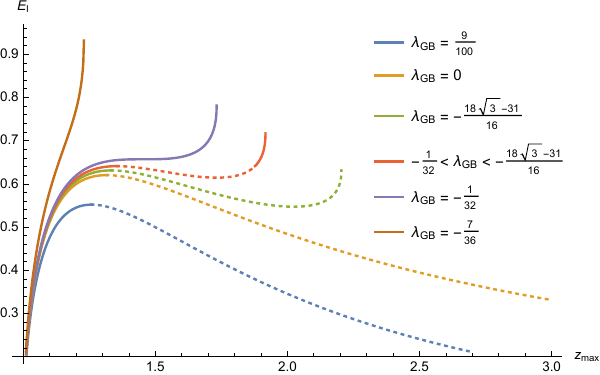}
\caption{Conserved quantity $E_\textrm{I}$ (\ref{island: EI1}) as a function of $z_{\max}$ for $d=4$.  We have $z_h\le z_{\max }\le z_s$ for negative $\lambda_{\text{GB}}$, where $z_h=1$ denotes the horizon and $z_s$ denotes the black hole singularity (\ref{islandzs}). The extremal surface is well-defined only for the $z_{\text{max}}$ in the solid lines.  As for the $z_{\text{max}}$ in the dotted lines, the entropy (\ref{island:HEEtypeI}) and time (\ref{island:timeI}) become complex. }
\label{Critical point}
\end{figure}

In this appendix, we analyze the time evolution of entanglement entropy related to the extremal surface passing through the horizon for various Gauss-Bonnet couplings. We find that there are three classes of conserved quantities $E_\textrm{I}$ (\ref{island: EI1}), which correspond to three different kinds of time evolutions of entanglement entropy. 

Let us first discuss the conserved quantity $E_\textrm{I}$ (\ref{island: EI1}), which plays an important role in studying the extremal surface passing through the horizon. As shown in  Fig.\ref{Critical point}, there are three kinds of  $E_\textrm{I}(z_{\text{max}})$ for different Gauss-Bonnet couplings.  In the normal case including Einstein gravity, (blue, orange and green lines of  Fig.\ref{Critical point}),  the local extremal point $\bar{z}_{\text{max}}$ is also the global maximum point for the conserved quantity (\ref{island: EI1}). In this case,  the entropy 
(\ref{island:HEEtypeI}) and time (\ref{island:timeI}) are monotonically increasing functions of $z_{\text{max}}$ for $z_h \le z_{\text{max}} \le \bar{z}_{\text{max}}$.  And we have $S(\bar{z}_{\text{max}})=t(\bar{z}_{\text{max}})=\infty$.  Note that we cannot select the range $ z_{\text{max}} > \bar{z}_{\text{max}} $,  otherwise the entropy (\ref{island:HEEtypeI}) and time (\ref{island:timeI}) become complex.  The calculations of entropy and time are  similar to those of Einstein gravity \cite{Chen:2020uac,Carmi:2017jqz}.  It turns out that the entropy increases with time. For simplicity, we do not repeat the calculations for the normal case. 

In the second case (red and purple lines of Fig.\ref{Critical point}), the local extremal point $\bar{z}_{\text{max}}$ (\ref{island:largetimezmax}) is no longer the global maximum point for the conserved quantity (\ref{island: EI1}). The critical case is labelled by the green line of Fig. \ref{Critical point}, which satisfies the relation
\begin{eqnarray}\label{appB:critical relation}
E_\text{I} (\bar{z}_{\max })  =E_\text{I} ({z}_{s}).
\end{eqnarray}
Recall that $\bar{z}_{\max }$ is given by (\ref{island:largetimezmax}), $z_s$ denotes the black hole singularity (\ref{islandzs}) and we have $z_h\le z_{\max }\le z_s$.  Solving (\ref{appB:critical relation}) together with (\ref{island: EI1},\ref{island:largetimezmax}, \ref{islandzs}), we get the critical Gauss-Bonnet coupling $\lambda_{f}$
\begin{eqnarray}\label{appB:lambdaGB1}
&&d=4: \quad  \lambda_{f}=- \frac{18\sqrt{3}-31}{16}\approx -0.011,\nonumber\\
&&d=5: \quad  \lambda_{f}\approx -0.017,\\
&&d=6: \quad  \lambda_{f}\approx -0.021.\nonumber
\end{eqnarray}

In the third case (dumpling line of Fig.\ref{Critical point}), there is no local extremal point $\bar{z}_{\text{max}}$ (\ref{island:largetimezmax}) for the conserved quantity (\ref{island: EI1}). The critical coupling between the second case and the third case is given by (\ref{island:conditionlambda})
\begin{eqnarray}\label{appB:lambdaGB2}
 \lambda_c=-\frac{(d-2)^2}{4d(3d-4)},
\end{eqnarray}
which is labelled by the purple line of Fig. \ref{Critical point}.

\begin{figure}[!ht]
  \centering
  \includegraphics[width=5cm]{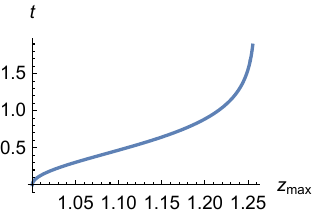}
  \includegraphics[width=5cm]{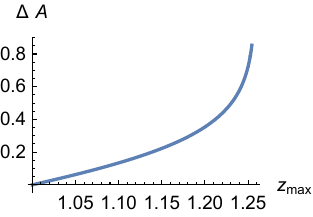}
  \includegraphics[width=4.5cm]{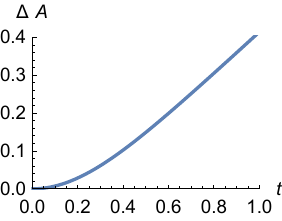}
  case I: $\lambda=9/100$  \\
  \includegraphics[width=5cm]{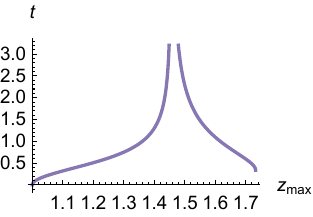}
  \includegraphics[width=5cm]{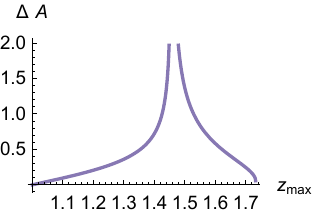}
  \includegraphics[width=4.5cm]{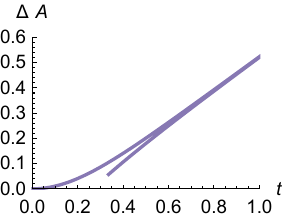}
  case II:  $\lambda=-1/32$  \\
  \includegraphics[width=5cm]{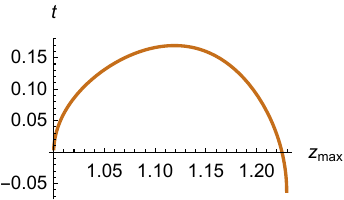}
  \includegraphics[width=5cm]{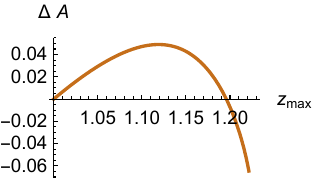}
  \includegraphics[width=4.5cm]{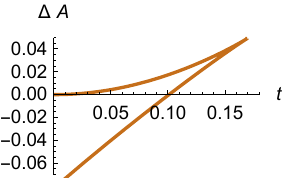}
  case III:  $\lambda=-7/36$  \\
\caption{Area and boundary time for extremal surfaces passing through the horizon for various couplings with $d=4$.}
\label{Atthreecases}
\end{figure}

Let us go on to study the time evolution of entanglement entropy for the above three cases. The normal case with 
\begin{eqnarray} \label{appB:case1}
{\text{case I}}:\  \lambda_f\le \lambda_{\text{GB}}\le \frac{(d-3) (d-2) \left(d^2-d+6\right)}{4 \left(d^2-3 d+6\right)^2},
\end{eqnarray}
is similar to Einstein gravity \cite{Chen:2020uac,Carmi:2017jqz}. For simplicity, we do not repeat the calculations. The conclusion is that the entropy increases with time and the growth rate approaches to (\ref{island:largetime1}) in the late time limit.  Note that the upper bound of (\ref{appB:case1}) comes from the causality constraint (\ref{GBconstraint}). 

As for the other two cases, one has to perform numeral calculations due to the complicated entropy functional  (\ref{island:HEEtypeI}).  To get more analytic understandings of two unusual cases, we simplify the entropy functional properly. Instead of applying the Jacobson-Myers formula (\ref{island:HEE}), we take the Bekenstein-Hawking entropy formula as a simplified toy model below \footnote{ We focus on Jacobson-Myers formula (\ref{island:HEE}) in the main text, the toy model of this appendix is just help to understand the qualitative behaviours of the time evolution of entropy.}
\begin{eqnarray} \label{Bekenstein_entropy}
S =\frac{\mathcal{A}}{4G_N}.
\end{eqnarray}
It does not change the qualitative behaviours of the time evolution of entropy for Gauss-Bonnet gravity, but simplifies the problem a lot.  Another advantage of this toy model is that it is similar to the complexity=volume conjecture of holographic complexity. The only difference is that the holographic entanglement entropy (\ref{Bekenstein_entropy}) is related to minimal codimension-two extremal surface, while the holographic complexity is related to maximum codimension-one extremal surface. Following the calculations of holographic complexity \cite{Carmi:2017jqz, An:2018dbz}, we obtain the expression of area $ \mathcal{A}$ and boundary time $t$,
\begin{eqnarray}
\mathcal{A} &=& V_{d-2}\int_0^{z_{\max}} \frac{dz}{z^{2(d-1)}\sqrt{E_{\mathcal{A}}(z_{\max})^2 - E_{\mathcal{A}}(z)^2}}  ,\label{A_of_z}\\
t &=&  \int_0^{z_{\max}} \frac{f_\infty E_{\mathcal{A}}(z_{\max})dz}{f(z)\sqrt{E_{\mathcal{A}}(z_{\max})^2-E_{\mathcal{A}}(z)^2}}, \label{t_of_z}
\end{eqnarray}
where $E_{\mathcal{A}}(z)^2=-\frac{f(z)}{f_\infty z^{2(d-1)}}$ and the conserved quantity of the toy model is given by
\begin{eqnarray} \label{toymodelEI}
E_{\mathcal{A}}(z_{\max})=z_{\max }^{1-d} \sqrt{-\frac{f\left(z_{\max }\right)}{f_{\infty }}},
\end{eqnarray}
which is the same as the conserved quantity of main text (\ref{island: EI1}) up to a constant factor.  For simplicity, we set $V_{d-2}=1$ below. Note that the extremal surface is well-defined only for the $z_{\text{max}}$ in the solid lines of Fig. \ref{Critical point}.  As for the $z_{\text{max}}$ in the dotted lines,  $E_{\mathcal{A}}(z_{\max})^2 - E_{\mathcal{A}}(z)^2$ can be negative for $0\le z \le z_{\text{max}}$, which yields complex area (\ref{A_of_z}) and time (\ref{t_of_z}).  Note also that, for any given $z_{\max} $, one get a unique time (\ref{t_of_z}). However, one time (\ref{t_of_z}) may correspond to two different $z_{\max}$ and thus two different areas (\ref{A_of_z}) for some couplings $\lambda_{\text{GB}}$. This is what happens for the two unusual cases.  And this leads to novel phase transitions of the entanglement entropy.

Now we are ready to discuss the time evolution of entanglement entropy for the second case (red and purple lines of Fig.\ref{Critical point})
\begin{eqnarray} \label{appB:case2}
{\text{case II}}:\ -\frac{(d-2)^2}{4d(3d-4)}=\lambda_c\le \lambda_{\text{GB}}< \lambda_f.
\end{eqnarray}
As we have mentioned above, one time (\ref{t_of_z}) corresponds to two different $z_{\max}$ and thus two different areas (\ref{A_of_z}) for case II. For example, we consider $\lambda_{\text{GB}}=\lambda_c=-1/32$, $z_h=1$ and $d=4$, which yields $\bar{z}_{\max}\approx 1.456$ and $z_s\approx 1.732$. Recall that $\bar{z}_{\max}$ is defined by (\ref{island:largetimezmax}) and $z_s$ denotes the singularity. Choosing $z_{\max 1}\approx 1.242$ and $z_{\max 2}\approx 1.699$, we obtain the same boundary time (\ref{t_of_z})
\begin{eqnarray} \label{boundarytimenumericalresult1}
&&t(z_{\max 1}) = \int_0^{z_{\max 1}}\frac{f_\infty E_{\mathcal{A}}(z_{\max 1})dz}{f(z)\sqrt{E_{\mathcal{A}}(z_{\max 1})^2-E_{\mathcal{A}}(z)^2}} \approx 0.596,\\
&&t(z_{\max 2}) = \int_0^{z_{\max 2}} \frac{f_\infty E_{\mathcal{A}}(z_{\max 2})dz}{f(z)\sqrt{E_{\mathcal{A}}(z_{\max 2})^2-E_{\mathcal{A}}(z)^2}} \approx 0.596,  \label{boundarytimenumericalresult2}
\end{eqnarray} 
but different extremal areas  (\ref{A_of_z})
\begin{eqnarray} \label{areanumericalresult1}
&&\Delta \mathcal{A}(z_{\max 1}) = \mathcal{A}(z_{\max 1}) - \mathcal{A}(z_h)\approx 0.262,\\
&&\Delta \mathcal{A}(z_{\max 2}) = \mathcal{A}(z_{\max 2}) - \mathcal{A}(z_h) \approx 0.248, \label{areanumericalresult2}
\end{eqnarray} 
where $\mathcal{A}(z_h)$ is the area at $t=0$.  
Note that we have $z_h<z_{\max 1}<\bar{z}_{\max}$ while $\bar{z}_{\max}<z_{\max 2}<z_s$. Note also that there is no range $\bar{z}_{\max}<z_{\max 2}$ in the usual case I. It is this new range $\bar{z}_{\max}< z_{\max 2}<z_s$ that makes the area $\mathcal{A}(t)$ non-single-valued. 
Now we have verified our claim that one time (\ref{t_of_z}) corresponds to two different extremal areas (\ref{A_of_z}) in case II. See Fig. \ref{Atthreecases} for the full time evolution of extremal areas. Note that we choose the minimal area for the entanglement entropy and this leads to zeroth-order phase transition in the time evolution of entanglement entropy.

Let us go on to study the time evolution of entanglement entropy for the third case (dumpling line of Fig.\ref{Critical point})
\begin{eqnarray} \label{appB:case3}
{\text{case III}}:\ -\frac{(d-2) (3 d+2)}{4 (d+2)^2}\le \lambda_{\text{GB}}< \lambda_c=-\frac{(d-2)^2}{4d(3d-4)},
\end{eqnarray}
where the lower bound is due to the causality constraint (\ref{GBconstraint}). 
Similar to case II, one time also corresponds to two different different extremal areas for case III, which yields similar zeroth-order phase transitions of entanglement entropy. The new feature for case III is that the extremal surface is well-defined only in a finite time, after that there is no extremal surface passing through the horizon. Let us explain how this happens. 

Note that the integrand in (\ref{t_of_z}) is divergent when $z=z_{\max}$, this can potentially but not necessarily lead to $t\to \infty$ so that the extremal surface can be defined at any late time. Note also that $z\sim z_{\max}$ is the only region that can contribute to potentially divergence of $t$. Below we focus on this region.


In case I and case II, we can take $E_{\mathcal{A}}'(z_{\max})=0$ for some values of $z_{\max}$. This leads to\begin{eqnarray} \label{Ezcase1}
&&E_{\mathcal{A}}(z)=E_{\mathcal{A}}(z_{\max})+\frac{1}{2} E_{\mathcal{A}}''(z_{\max}) (z_{\max}-z)^2+... \nonumber\\
&& E_{\mathcal{A}}(z_{\max})^2-E_{\mathcal{A}}(z)^2=-E_{\mathcal{A}}(z_{\max})E_{\mathcal{A}}''(z_{\max}) (z_{\max}-z)^2+...
\end{eqnarray} 
From (\ref{t_of_z}, \ref{Ezcase1}), we have 
\begin{eqnarray} \label{boundarytime1}
t\sim \int^{z_{\max}} \frac{dz}{z_{\max}-z}\sim  \lim_{z\to z_{\max}}-
\ln(z_{\max}-z) \to \infty,
\end{eqnarray} 
 which means that the extremal surface is well-defined for $t\to \infty$. 
 
 On the other hand, in the case III, we have $E_{\mathcal{A}}'(z_{\max})\ne 0$ for all values of $z_{\max}$. See the dumpling line of Fig. \ref{Critical point}. Then we have  
 \begin{eqnarray} \label{Ezcase2}
&&E_{\mathcal{A}}(z)=E_{\mathcal{A}}(z_{\max})-E_{\mathcal{A}}'(z_{\max}) (z_{\max}-z)+... \nonumber\\
&& E_{\mathcal{A}}(z_{\max})^2-E_{\mathcal{A}}(z)^2=2E_{\mathcal{A}}(z_{\max})E_{\mathcal{A}}'(z_{\max}) (z_{\max}-z)+...
\end{eqnarray} 
which yields around $z\sim z_{\max}$ that
\begin{eqnarray} \label{boundarytime2}
t\sim \int^{z_{\max}} \frac{dz}{\sqrt{z_{\max}-z}}\sim \lim_{z\to z_{\max}} -\sqrt{z_{\max}-z}=0.
\end{eqnarray} 
This means that the potentially divergence around $z\sim z_{\max}$ disappears and the time become finite.  In other words, the extremal surface can not be defined after some finite time for sufficiently negative $\lambda_\text{GB}$. 

To summarize, we have two critical couplings $\lambda_{f}$ and $\lambda_{c}$  for Gauss-Bonnet gravity and the two critical couplings $\lambda_{f}$ and $\lambda_{c}$ lead to three kinds of time evolution of entanglement entropy relevant to the extremal surfaces passing through the horizon.  See Fig. \ref{Atthreecases} for the toy model and Fig. \ref{entropytypeII} for real case. In the normal case with (\ref{appB:case1}), the entropy increases with time and the growth rate approaches to (\ref{island:largetime1}) in the late time limit. In the second case with (\ref{appB:case2}),  the entropy firstly increases, then drops suddenly and continues to increase until the growth rate approaches to (\ref{island:largetime1}). Interestingly, there is a zeroth-order phase transition of entanglement entropy at early times. In the third case with 
(\ref{appB:case3}), there is also  a zeroth-order phase transition at early times.  However, the extremal surfaces passing through the horizon disappears after some finite time. This is problematic in AdS/CFT, but not a problem in AdS/BCFT. Due to the appearance of island, the entanglement entropy is well-defined in any late time in AdS/BCFT.

\end{document}